\documentclass[usegraphicx,usenatbib,useAMS]{mn2e}

\usepackage{multirow}
\usepackage{xcolor}
\usepackage{aas_macros}
\usepackage{amssymb}
\usepackage{xspace}
\usepackage[normalem]{ulem}
\usepackage{enumitem}
\usepackage{amsmath}

\newcommand{\lsim}{\lesssim} 

\newcommand{\kms}{\,{\rm km}\,{\rm s}^{-1}}
\newcommand{\Msol}{\,{\rm M}_{\odot}}

\newcommand{\Mpc}{\,{\rm Mpc}}

\newcommand{\GALFORM}{\textsc{galform}\xspace}
\newcommand{\lgalaxy}{\textsc{l-galaxies}\xspace}
\newcommand{\MORGANA}{\textsc{morgana}\xspace}

\newcommand{\AREPO}{\textsc{arepo}\xspace}
\newcommand{\SUBFIND}{\textsc{subfind}\xspace}

\newcommand{\SAG}{\textsc{sag}\xspace}
\newcommand{\SAGE}{\textsc{sage}\xspace}
\newcommand{\DARKSAGE}{\textsc{dark sage}\xspace}
\newcommand{\SHARK}{\textsc{shark}\xspace}
\newcommand{\MERAXES}{\textsc{meraxes}\xspace}

\newcommand{\eagle}{\textsc{eagle}\xspace}

\title[Cooling angular momentum model comparison]
{How well is angular momentum accretion modelled in semi-analytic galaxy formation models?}

\author[Hou et al.]  {Jun Hou\thanks{\scriptsize E-mail:
    jun.hou@durham.ac.uk (JH);cedric.lacey@durham.ac.uk (GCL)},$^{1,2}$ Cedric G. Lacey\footnotemark[1],$^2$
  Carlos. S. Frenk$^2$
  \\
  $^{1}$Shanghai Key Lab for Astrophysics, Shanghai Normal University, 
  100 Guilin Road, Shanghai 200234, China
  \\
  $^{2}$Institute for Computational Cosmology, Department of Physics,
  University of Durham, South Road, Durham, DH1 3LE, UK}

\begin{document}

\maketitle

\begin{abstract} Gas cooling and accretion in haloes delivers mass and angular momentum onto galaxies. In this work, we investigate the accuracy of the modelling of this important process in several different semi-analytic (SA) galaxy formation models (\GALFORM, \lgalaxy and \MORGANA) through comparisons with a hydrodynamical simulation performed with the moving-mesh code \AREPO. Both SA models and the simulation were run without any feedback or metal enrichment, in order to focus on the cooling and accretion process. All of the SA models considered here assume that gas cools from a spherical halo. We found that the assumption that the gas conserves its angular momentum when moving from the virial radius, $r_{\rm vir}$, to the central region of the halo, $r \sim 0.1 r_{\rm vir}$, is approximately consistent with the results from our simulation, in which gas typically retains $70-80\%$ of its angular momentum during this process. We also found that, compared to the simulation, the \MORGANA model tends to overestimate the mean specific angular momentum of cooled-down gas, the \lgalaxy model also tends to overestimate this in low-redshift massive haloes, while the two older \GALFORM models tend to underestimate the angular momentum. In general, the predictions of the new \GALFORM cooling model developed by \citeauthor{new_cool} agree the best with the simulation. 
\end{abstract}

\begin{keywords}
methods: numerical -- galaxies: evolution -- galaxies:formation
\end{keywords}

\section{Introduction}\label{sec:introduction}
The cooling of gas in haloes and its accretion onto galaxies not only brings mass into galaxies but also angular momentum. This process, which we refer to as angular momentum accretion, is a major channel for galaxies to gain angular momentum, and therefore plays an important role in determining galaxy sizes and further, star formation rates and gas fractions in galaxies. It should be included in any theoretical model of galaxy formation based on consideration of the underlying physical processes.

There are two main classes of physically-based theoretical galaxy formation models, namely hydrodynamical simulations and semi-analytic (SA) models. The former \citep[e.g.][]{illustris_simulation,eagle_simulation} try to numerically solve the hydrodynamical equations involved in galaxy formation. This provides very rich details of the formation and properties of galaxies, but also results in a high computational cost, for which generating a large galaxy sample for statistical studies remains challenging. In contrast, SA models \citep[e.g.][]{white_rees_1978,WF1991,baugh_2006_SA_review,benson_2010_SA_review} adopt simple analytic recipes to model various physical processes in galaxy formation, and mainly focus on global galaxy properties such as the total stellar mass and total cold gas mass in a galaxy. This significantly reduces the computational cost and makes SA models a good complement to hydrodynamical simulations. Combining these two methods may provide the most efficient strategy for improving our current theoretical understanding of galaxy formation. To achieve this goal, the recipes in an SA model should be as physically realistic as possible.

The modelling of angular momentum accretion in current SA models covers a wide range of sophistication. A brief review is given below.

The simplest approach does not explicitly follow the angular momentum associated with cooling gas flows, but simply assumes that at any given time the mean specific angular momentum of a galaxy is the same as that of its host halo. This is the approach followed in the \MORGANA SA model \citep{morgana1,morgana2}. Some other SA models, e.g.\ \SAG\citep{SAG_model_2018}, \SAGE\citep{SAGE_model_2016} and \MERAXES\citep{MERAXES_model_2016}, also adopt similar modelling, and they will be further discussed in \S\ref{sec:morgana_model}.

The \lgalaxy model \citep[e.g.][]{munich_model1,munich_model_Guo11} is more sophisticated. It calculates the angular momentum of gas accretion by assuming that the mean specific angular momentum of the cooled-down gas that is accreted onto a galaxy within a given timestep is the same as that of the galaxy's host halo at that time. There are also other SA models that adopt similar assumptions, e.g.\ \SHARK\citep{SHARK_model_2018} and \DARKSAGE\citep{Dark_SAGE_model_2016}, and more detailed discussions are given in \S\ref{sec:lgalaxies_model}.

The \GALFORM model \citep[e.g.][]{cole2000,galform_bower2006,galform_gonzalez2014,galform_lacey2015} is even more sophisticated. It first follows the evolution of the angular momentum of the halo gas, which is the gas in dark matter haloes but not in galaxies, by assuming that the gas and dark matter accreted onto a halo have the same specific angular momentum. Then it calculates the angular momentum of the gas that cools and accretes onto a galaxy via an assumed specific angular momentum distribution for the halo gas. In older \GALFORM cooling models \citep{cole2000,galform_bower2006,benson_bower_2010_cooling}, it is assumed $\boldsymbol{j}(r)\propto r$, where $\boldsymbol{j}(r)$ is the specific angular momentum of a spherical halo gas shell with radius $r$. This form was motivated by hydrodynamical simulations without radiative cooling. In the new \GALFORM cooling model introduced in \citet{new_cool}, the evolution of $\boldsymbol{j}(r)$ induced by cooling and accretion is also taken into account.

In this work, we assess the accuracy of all of these SA models for angular momentum accretion by comparing with the results of high resolution cosmological hydrodynamical simulations. Since recent simulations are largely able to resolve the gas cooling process and associated angular momentum accretion, these comparisons should provide a good test for SA models. Previous works comparing SA cooling models with hydrodynamical simulations \citep[e.g.][]{Benson_2001_comp,Yoshida_2002_comp,Helly_2003_comp,Cattaneo_2007_comp,morgana2,Saro_2010_comp,Lu_2011_comp,Hirschmann_2012_comp,monaco_2014_comp, hou_cooling_comp} mainly focused on the gas mass transported into galaxies, and paid little attention to the associated angular momentum flow.

In \citet{hou_cooling_comp} (hereafter Paper I), the mass cooling and accretion predictions of the above-mentioned SA models were compared with results from cosmological hydrodynamical simulations. It was found that in general the predictions of the new \GALFORM cooling model agree well with simulations, and this agreement is the best among all of the SA models considered. It is therefore of great interest to investigate the accuracy of this and other SA cooling models in predicting angular momentum accretion.

There have been some previous investigations of angular momentum accretion using hydrodynamical simulations, and a few of these works also included a brief comparison with SA models \citep[e.g.][]{Stewart_j_2013,Danovich_j_2015,Stevens_cooling_2017}. However, those works only considered the simplest SA models of angular momentum accretion, rather than covering the whole range of SA models. Also, many of the above works (e.g.\ \citeauthor{Stewart_j_2013} and \citeauthor{Danovich_j_2015}), mainly focused on the regime where cold filamentary gas is an important component of the halo gas, while paying little attention to the regime where a hot spherical gas halo dominates the halo gas. In this work we consider both of these regimes.

In this work, as also in many of those mass accretion comparison papers cited above, we turn off all feedback and metal enrichment processes. The current work and those previous works together should provide a complete benchmark for SA cooling models in this simplified situation. If a model gives predictions with acceptable accuracy in this simplified situation, but shows large deviations when feedback is included, then it is clear that the parts involving feedback in this model are not accurate and need to be improved. In contrast, if a model is clearly biased in the situation without feedback, but nevertheless gives reasonable results when a specific feedback model is added, then this model probably involves some assumptions that are only valid for specific situations or parameters related to feedback, and therefore has a small dynamic range for its predictions. This small range would limit the information on galaxy formation physics that one could extract through studying the model behaviour in the whole parameter space.

A more realistic treatment of galaxy formation of course includes feedback. But to compare angular momentum predictions from SA models and simulations in this more complex situation, one should ensure that the differences seen between the two methods are not caused by differences in their modelling of feedback. This requires that SA models and simulations adopt physically identical modelling of feedback, which is both complex and difficult to achieve, and we defer this to future work.

This paper is organized as follows. \S\ref{sec:method_simulation} introduces the hydrodynamical simulations used in this work, and \S\ref{sec:method_j_measurement} describes how the angular momentum of gas accretion onto galaxies is measured in these simulations, while \S\ref{sec:method_SA_j_model} provides a more detailed description of the SA angular momentum accretion models considered in this work. Our main results are given in \S\ref{sec:results}, with \S\ref{sec:assumption_check} checking various assumptions involved in SA calculations of angular momentum accretion and \S\ref{sec:comparison} comparing the mean specific angular momenta of the cooled-down gas predicted by SA models and the hydrodynamical simulation. We give a summary of our results in \S\ref{sec:summary}.

\section{Methods} 
\label{sec:methods}

\subsection{Simulations}
\label{sec:method_simulation}
The simulations in this work were performed using the moving-mesh code \AREPO \citep{arepo}. Compared to the widely used SPH (Smooth Particle Hydrodynamics) method, the moving-mesh method can largely avoid artificial shock broadening and turbulence damping, to which the cooling calculation is sensitive \citep[e.g.][]{Bauer_2012,Nelson2013_hydro_cooling}. But note that in the moving-mesh method, averaging quantities within cells still results in some numerical diffusion effects.

In a grid-based simulation, it is difficult to trace Lagragian gas motion, especially without using tracer particle techniques \citep[e.g.][]{arepo_trace_particle}. This becomes a limitation when we try to check various assumptions in SA angular momentum calculations in \S\ref{sec:assumption_check}. There we adopt a simple but rough method to approximately trace Lagragian motion. However, our primary comparison in this work uses the measurement of cooled-down angular momentum and does not rely on tracing Lagragian gas motion, so the comparison of predictions of this quantity from SA models and simulations (in \S\ref{sec:comparison}) is unaffected by this limitation.

We use the simulation suites described in Paper~I, and more details can be found there. These simulations were run in two cubes with comoving sizes $50\Mpc$ and $25\Mpc$ respectively. There are $752^3$ and $376^3$ dark matter particles in the large and small cubes respectively, both corresponding to a dark matter particle mass $9.2\times 10^6\Msol$. For the hydrodynamical simulations, there are initially the same number of gas cells as dark matter particles. We also performed dark matter only simulations in order to construct halo merger trees for the SA models. Data from the simulations was output at $128$ snapshots, which are evenly spaced in $\log(1+z)$, from $z=19$ to $z=0$, where $z$ is the redshift. The physical time interval between two adjacent snapshots is approximately $0.25t_{\rm dyn}$, with $t_{\rm dyn}=r_{\rm vir}/V_{\rm vir}$ being the halo dynamical timescale, $r_{\rm vir}$ the halo virial radius and $V_{\rm vir}$ the halo virial velocity. All the simulations were run with the WMAP-7 cosmological parameters \citep{CMB_obs_WMAP7}: $\Omega_{\rm m0}=0.2726$, $\Omega_{\rm \Lambda 0}=0.7274$, $\Omega_{\rm b0}=0.0455$, $H_{\rm 0}=70.4\kms{\Mpc}^{-1}$, and an initial power spectrum with slope $n_{\rm s}=0.967$ and normalization $\sigma_{\rm 8}=0.810$. In this work, we use the hydrodynamical and dark matter only simulations in the $50\Mpc$ cube for our main analysis, while use the hydrodynamical simulation in the $25\Mpc$ cube for test purposes.

The structures formed in the simulations are first identified using the friends-of-friends (FOF) algorithm \citep{FOF_algorithm}, and then each FOF group is split into subgroups using \SUBFIND \citep{munich_model1}.
Halo merger trees are then constructed using the Dhalo algorithm \citep{Dhalo_tree1,Dhalo_tree2}, in which subgroups identified by \SUBFIND at different snapshots are linked by matching their most bound dark matter particles, and subgroups at the same snapshot are grouped into Dhalos by examing their separations. Unless otherwise stated, all halo masses used in this paper are masses of Dhalos. The merger trees in the hydrodynamical simulation and dark matter only simulation are built separately and then linked by cross matching the $50$ most bound dark matter particles of the base node at $z=0$. Two linked trees are treated as the merger trees of the same halo in different simulations.

In the hydrodynamical simulations, a gas cell is allowed to cool only when its temperature is above the following threshold
\begin{equation}
T_{\rm cool,lim}=3.5\times 10^4\times [\Omega_{\rm m0}(1+z)^3+\Omega_{\rm \Lambda 0}]^{1/3}\,{\rm K} .
\label{eq:T_lim_cool}
\end{equation}
$T_{\rm cool,lim}$ corresponds approximately to the virial temperature of a halo with mass $2\times 10^{10}\Msol$. This temperature threshold restricts the gas cooling to occur only in well-resolved haloes (resolved with at least $2000$ dark matter particles). 

Gas is turned into stars once its density is higher than $\delta_{\rm str,lim}\bar{\rho}_{\rm gas}$ and its temperature is lower than $T_{\rm str,lim}$. Here $\bar{\rho}_{\rm gas}={\rm \Omega}_{\rm b}(z)\rho_{\rm crit}(z)$ is the mean gas density, with ${\rm \Omega}_{\rm b}(z)$ and $\rho_{\rm crit}(z)$ the baryon fraction and critical density at redshift $z$ respectively, and $\delta_{\rm str,lim}$ and $T_{\rm str,lim}$ are two parameters. We adopt $\delta_{\rm str,lim}=10^4$ and $T_{\rm str,lim}=\min[10^5\,{\rm K},\ T_{\rm cool,lim}]$. This recipe rapidly turns the gas into collisionless stellar particles once it cools down and is accreted by galaxies, the purpose of this being to save computation time. More details of the cooling calculation in these simulations are given in Paper~I. Note that in order to focus on the gas cooling and accretion processes, we turn off any feedback and metal enrichment processes in the simulations.

\subsection{Measuring angular momenta in hydrodynamical simulations}
\label{sec:method_j_measurement}
In this work we focus on the gas cooling and associated angular momentum accretion onto central galaxies. The central galaxy of a Dhalo is defined as that associated to the most massive subgroup of the Dhalo. As mentioned above and described in further detail in Paper~I, the gas cooled down and accreted onto a central galaxy is rapidly turned into stellar particles, so that the gas accreted between two adjacent snapshots approximately corresponds to the mass of stellar particles newly formed in this galaxy between these snapshots (gas cooling onto satellites does not last for long, so the satellites merging with a central galaxy are almost all gas-poor, and so any gas they bring in provides only minor contributions to the newly formed stellar particles). Therefore, to measure the mass of the gas that cools down between snapshot $i$ and snapshot $i+1$ (with snapshot $i+1$ corresponding to lower redshift), we just need to measure the total mass of the stellar particles that form between snapshot $i$ and snapshot $i+1$.

On the other hand, the measurement of the angular momentum brought in by the cooled-down gas is more complicated. This is because once the gas is accreted onto a central galaxy, its angular momentum may be redistributed within the galaxy by gravitational torques and hydrodynamical interactions with existing gas. The timescale for this redistribution is expected to be a few galaxy dynamical timescales, $t_{\rm dyn,gal}=r_{\rm gal}/V_{\rm gal}$, with $r_{\rm gal}$ and $V_{\rm gal}$ respectively the radius and circular velocity of this galaxy \citep{Danovich_j_2015}. Noting that $V_{\rm gal}\sim V_{\rm vir}$ and $r_{\rm gal}\sim 0.1r_{\rm vir}$, one then has $t_{\rm dyn,gal}\sim 0.1t_{\rm dyn}$, with $t_{\rm dyn}=r_{\rm vir}/V_{\rm vir}$ the halo dynamical timescale. The time intervals between adjacent snapshots in our simulations are around $0.25t_{\rm dyn}$, within which significant angular momentum redistribution could occur. Therefore, the angular momentum of the newly formed stellar particles measured at snapshot $i+1$ need not represent well the original angular momentum associated with the cooled-down gas.

To remove the effects of the above-mentioned angular momentum redistribution within a galaxy on our measurement of the accreted angular momentum, rather than considering the angular momentum of the newly formed stellar particles, we consider the angular momentum change of a system from snapshot $i$ to $i+1$. At snapshot $i+1$, the system under consideration is just the central galaxy, while at snapshot $i$, it consists of the central galaxy and the satellites that are going to merge with it between snapshots $i$ and $i+1$. We first select at snapshot $i+1$ all stellar particles within an aperture (defined later) around the centre of the most massive subgroup of a given Dhalo, and calculate its total angular momentum, $\boldsymbol{J}_{\rm i+1}$, at this snapshot. Then we identify in this collection of particles all of the stellar particles that also exist at snapshot $i$ (some of these particles may be in merging satellites at snapshot $i$), and calculate the total angular momentum of those particles, $\boldsymbol{J}_{\rm i}$, at snapshot $i$. We then assume that the angular momentum brought in by the accreted cooled-down gas between snapshots $i$ and $i+1$ is given by $\Delta \boldsymbol{J}_{\rm cool,i+1}=\Delta \boldsymbol{J}_{\rm i+1}\equiv\boldsymbol{J}_{\rm i+1}-\boldsymbol{J}_{\rm i}$. Both $\boldsymbol{J}_{\rm i+1}$ and $\boldsymbol{J}_{\rm i}$ are calculated in the centre-of-mass frames of the corresponding stellar particle groups, and the reference points for the angular momenta are the corresponding centres of mass. Note that the angular momentum at snapshot $i$ includes the orbital angular momentum of merging satellite galaxies. If we assume that the effect of external torques on the stellar particles during this time interval can be neglected, then this orbital angular momentum all ends up as part of the total angular momentum at snapshot $i+1$. In that case, the change in angular momentum $\Delta \boldsymbol{J}_{\rm i+1}$ is entirely due to gas cooling. This assumption is discussed further below.

The aperture that we used is formed as follows. We first select all stellar particles within $50\,{\rm comoving\ kpc}$ from the centre of the most massive subgroup of a given Dhalo. We then derive the $90$ percentile of the distances of these stellar particles from the subgroup centre, $r_{\rm 90}$, and set the aperture size to be $2r_{\rm 90}$. We use this variable aperture instead of a simple fixed-size aperture because the latter sometimes includes too many stellar particles bound to the subgroup but not to the central galaxy (such as halo stars). These particles typically have a large angular momentum change between two snapshots due to the gravitational potential of the halo, and mask the angular momentum change induced by gas cooling. In Appendix~\ref{app:aperture_size} we compare the measurements done with aperture radius $2r_{\rm 90}$ with those done with aperture radius $1.1r_{\rm 90}$ and $3r_{\rm 90}$ respectively, and find that the differences caused by varying aperture size is smaller than the differences between the SA models and the simulation seen in Fig.~\ref{fig:cool_vJ_comp}, so the results in Fig.~\ref{fig:cool_vJ_comp} is not sensitive to the choice of aperture size.

\begin{figure*}
    \centering
    \includegraphics[width=1.0\textwidth]{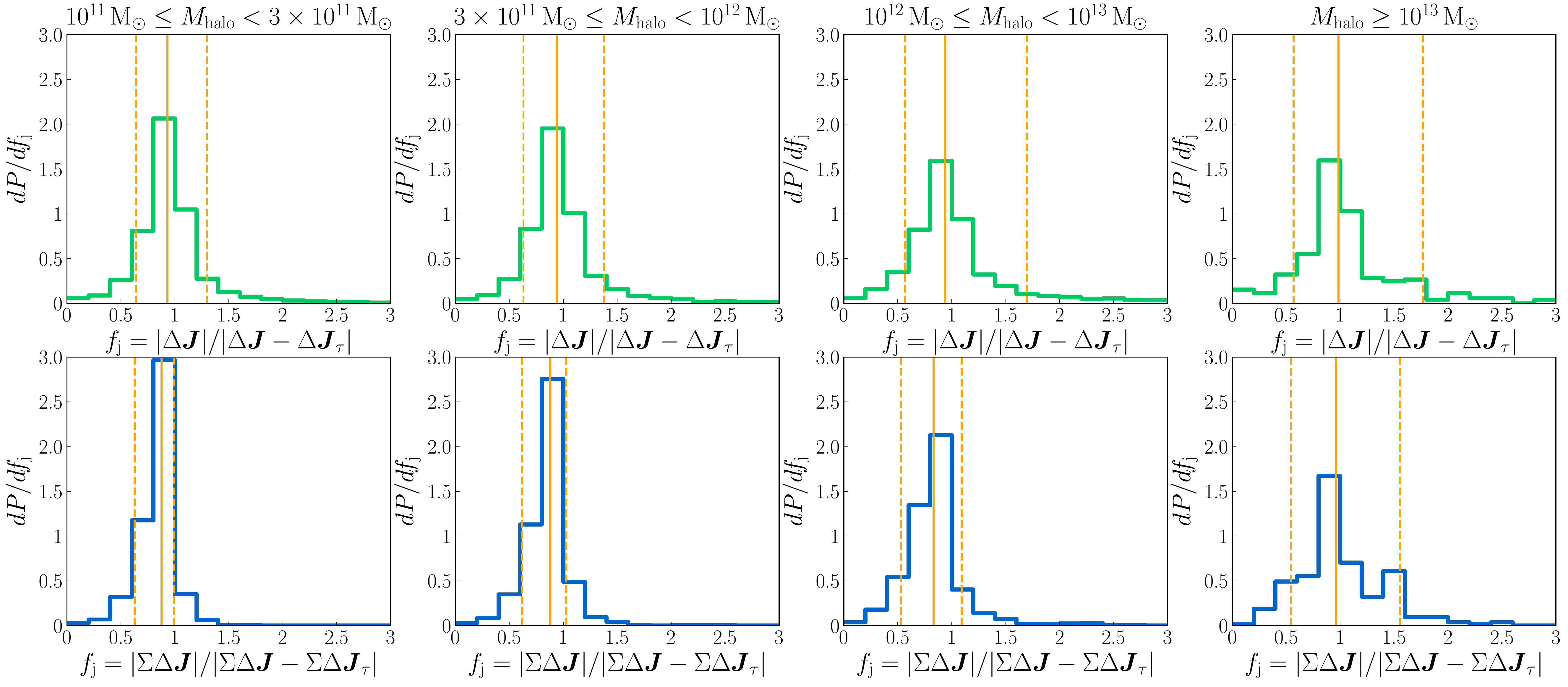}
    \includegraphics[width=1.0\textwidth]{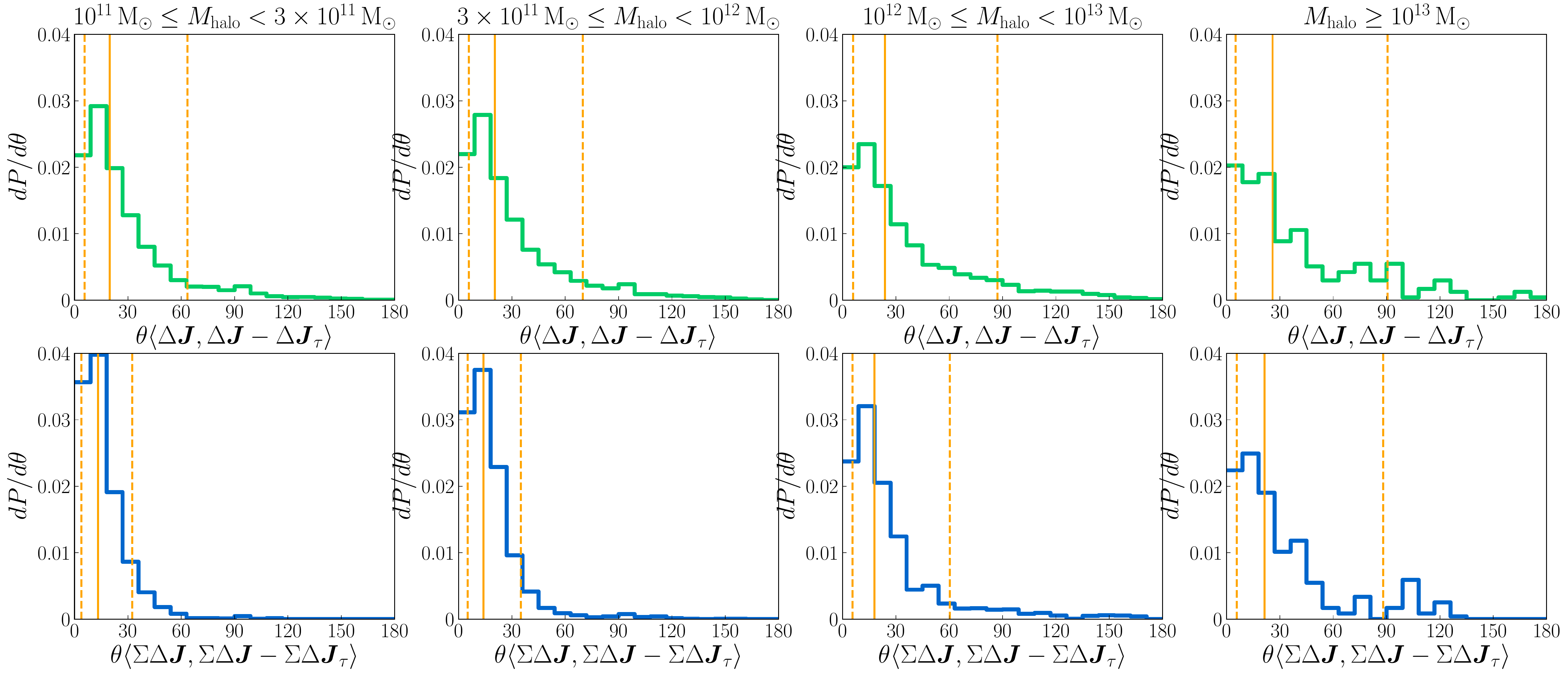}
    \caption{Relative importance of the angular momentum change induced by gravitational torques. Here $\Delta\boldsymbol{J}_{\rm \tau}$ is the estimation of the angular momentum change induced by gravitational torques from outside a given system (defined in \S\ref{sec:method_j_measurement}) between two adjacent snapshots, and $\Delta\boldsymbol{J}$ is the total angular momentum change of this system during the same time interval, while $\Sigma\Delta\boldsymbol{J}_{\rm \tau}$ and $\Sigma\Delta\boldsymbol{J}$ are the corresponding cumulative changes. Only haloes along the main branch of a given merger tree are included. Haloes are grouped into four ranges of $z=0$ descendant mass. Each column is for a halo sample, with the $z=0$ descendant halo mass range given at the top of that column. The first row gives the distribution of the ratios of magnitudes of $\Delta\boldsymbol{J}$ and $\Delta\boldsymbol{J}-\Delta\boldsymbol{J}_{\rm \tau}$, and the third row gives the distribution of angular offsets between them. The second and fourth rows give the distributions of ratios of magnitude and angular offsets of the corresponding cumulative quantities, $\Sigma\Delta\boldsymbol{J}$ and $\Sigma\Delta\boldsymbol{J}-\Sigma\Delta\boldsymbol{J}_{\rm \tau}$. In each panel, the vertical solid line indicates the median while the vertical dashed lines show the 10 and 90 percentiles. The contributions to these distributions from snapshots at which massive satellite galaxy mergers (with mass ratio greater than $0.1$) have occurred have been removed. }
    
    \label{fig:torque_dJ}
\end{figure*}

Apart from the angular momentum brought in by the cooled-down gas, the central galaxy can also change its angular momentum due to gravitational torques from outside the galaxy, which would therefore contribute to $\Delta \boldsymbol{J}_{\rm i+1}$. To estimate the importance of this contribution, we calculated the gravitational torque on each stellar particle in the $25\Mpc$ cube simulation, and estimate the torque induced angular momentum change as $\Delta \boldsymbol{J}_{\rm \tau}=(t_{\rm i+1}-t_{\rm i})\boldsymbol{\tau}_{\rm tot}$, where $t_{\rm i+1}$ and $t_{\rm i}$ are the age of universe at snapshots $i+1$ and $i$ respectively, and $\boldsymbol{\tau}_{\rm tot}$ is the total gravitational torque on the selected stellar particles at snapshot $i+1$. (Note that the contribution of internal torques cancels in $\boldsymbol{\tau}_{\rm tot}$, so that what is measured is the torque due to particles outside the galaxy at snapshot $i+1$.) 

This estimate of the angular momentum change due to torques is however only rough, as the torques should be integrated over time to derive angular momentum changes, while $\boldsymbol{\tau}_{\rm tot}(t)$ is only sampled at discrete times defined by the output snapshots. On the other hand, the total angular momentum change $\Delta\boldsymbol{J}_{\rm i+1}$ is calculated from the positions and velocities of stellar particles, which are evolved through simulation internal timesteps that are much finer than the snapshot intervals. Therefore, while $\Delta\boldsymbol{J}_{\rm \tau}$ as calculated here provides an estimate of the size of gravitational torque effects, it cannot be used to accurately subtract this torque contribution from $\Delta\boldsymbol{J}_{\rm i+1}$. In Appendix~\ref{app:torque_subtraction}, we test the effects of making this correction for the 25~Mpc cube, and find that the results of comparing SA predictions to simulation data only slightly change if we use $\Delta\boldsymbol{J}_{\rm i+1}-\Delta\boldsymbol{J}_{\rm \tau}$ in place of $\Delta\boldsymbol{J}_{\rm i+1}$. For both of these reasons, we do not calculate gravitational torques for our $50\Mpc$ cube simulation, and derive simulation results using $\Delta\boldsymbol{J}_{\rm i+1}$, i.e. the angular momentum change uncorrected for the gravitational torque contribution.

Fig.~\ref{fig:torque_dJ} compares $\Delta\boldsymbol{J}_{\rm i+1}$ with $\Delta\boldsymbol{J}_{\rm i+1}-\Delta\boldsymbol{J}_{\rm \tau}$. The latter quantity is approximately corrected for external gravitational torque effects. The medians of magnitude ratio of these two quantities in all four halo mass bins are all close to $1$, with 10-90 percentiles about $[0.5, 1.5]$. The median angular offset is about $30^{\circ}$ with corresponding 10-90 percentile widening from around $[0^{\circ}, 60^{\circ}]$ to around $[0^{\circ}, 90^{\circ}]$ when moving from low mass to high mass haloes. The width of these percentiles indicates that the external torque contamination in $\Delta \boldsymbol{J}_{i+1}$ is not completely negligible, so the derived evolution of $\Delta \boldsymbol{J}_{\rm cool,i}$ along a given branch of a halo merger tree could contain significant noise due to external torques. However, we find that the cumulative angular momentum change, $\boldsymbol{J}_{\rm cool}(<t_{\rm i})$, which is defined as 
\begin{equation}
\boldsymbol{J}_{\rm cool}(<t_{\rm i})=\sum_{j=i_{\rm start}}^{i} \Delta \boldsymbol{J}_{\rm cool,j}, \label{eq:J_cool_def}
\end{equation} 
(with $i_{\rm start}$ the index of the snapshot at which cooling started) is less affected and thus more robust. Fig.~\ref{fig:torque_dJ} also compares the cumulative quantities $\Sigma\Delta\boldsymbol{J}_{\rm i+1}$ and $\Sigma\Delta\boldsymbol{J}_{\rm i+1}-\Sigma\Delta\boldsymbol{J}_{\rm \tau}$ for different snapshots, where $\Sigma\Delta\boldsymbol{J}_{\rm \tau}$ is an approximate correction for effects of external gravitational torques, and we use $\Sigma\Delta\boldsymbol{J}_{\rm i+1}$ to mean $\boldsymbol{J}_{\rm cool}(<t_{\rm i+1})$ for this figure. It can be seen that for this pair of cumulative quantities, the median magnitude ratios are still close to $1$, and the median angular offsets are still around $30^{\circ}$, but the 10-90 percentiles ranges are narrower (the reduction of 10-90 percentile width is not very obvious for angular offsets in the halo mass bin with $M_{\rm halo}>10^{13}\Msol$, but it can still be seen that the distribution is concentrated towards $0^{\circ}$), and the situation is indeed better than for $\Delta\boldsymbol{J}_{\rm cool,i}$. For the cumulative angular momentum change, there is still scatter caused by external gravitational torque effects, as can be seen from Fig.~\ref{fig:torque_dJ}, but it is small compared to the differences between the predictions of the SA models and the simulation discussed in \S\ref{sec:comparison}. The overall conclusions of the comparison in \S\ref{sec:comparison} are therefore not significantly affected.

The medians of the ratio $|\Sigma\Delta\boldsymbol{J}_{\rm i+1}|/|\Sigma\Delta\boldsymbol{J}_{\rm i+1}-\Sigma\Delta\boldsymbol{J}_{\rm \tau}|$ are below $1$, indicating that the torques tend to be anti-aligned with the accreted angular momentum. We found that the median angles between $\Sigma\Delta\boldsymbol{J}_{\rm \tau}$ and $\Sigma\Delta\boldsymbol{J}_{\rm i+1}-\Sigma\Delta\boldsymbol{J}_{\rm \tau}$ are around $120^{\circ}$ for the above-mentioned four halo mass bins (with a 10-90 percentile range around $90-160^{\circ}$), supporting this tendency for anti-alignment.

In this work we only consider the cumulative angular momentum change $\boldsymbol{J}_{\rm cool}(<t_{\rm i})$, because it is less contaminated by torques. Here the summation in equation~(\ref{eq:J_cool_def}) is limited to the main branch of a given halo merger tree, i.e.\ formed by finding the most massive progenitor at each preceding snapshot. Further, the specific cumulative angular momentum is
\begin{equation}
\boldsymbol{j}_{\rm cool}(<t_{\rm i})=\frac{\boldsymbol{J}_{\rm cool}(<t_{\rm i})}{M_{\rm cool}(<t_{\rm i})}, \label{eq:j_cool_def}
\end{equation}
where $M_{\rm cool}(<t_{\rm i})$ is the cumulative mass cooled down by time $t_{\rm i}$, and is calculated by adding the masses of stellar particles newly formed at each snapshot before $t_{\rm i}$ along the main branch of the merger tree. 

One source of gravitational torques from outside the system is the dynamical friction force exerted on merging satellites by the dark matter halo. When the satellite to central mass ratio is high, strong dynamical friction leads to rapid decay of the satellite's orbit, and during the time interval between two snapshots, which is around $0.25t_{\rm dyn}$, significant orbital angular momentum loss can be induced. This angular momentum loss would be included in the angular momentum change, $\Delta \boldsymbol{J}_{\rm i+1}$, mentioned above, and can strongly affect even cumulative quantities like $\boldsymbol{J}_{\rm cool}(<t_{\rm i+1})$ and $\boldsymbol{j}_{\rm cool}(<t_{\rm i+1})$. Without information on the satellite motion between two adjacent snapshots, the effects of strong dynamical friction on the angular momentum change of the central galaxy cannot be easily corrected for, and therefore we choose to remove the contributions from snapshots having massive satellite mergers. More specifically, in the summation in equation~(\ref{eq:J_cool_def}) we remove the contributions $\Delta J_{\rm cool,j}$ from timesteps during which galaxy mergers occur with satellite to central stellar mass ratios larger than $0.1$. In total this removes about $8$ percent snapshots for merger trees with $M_{\rm halo}(z=0)\geq 10^{13}\Msol$, while the fraction is smaller for lower halo masses. Correspondingly, we also remove the contributions of these snapshots to $M_{\rm cool}(<t_{\rm i})$ in equation~(\ref{eq:j_cool_def}).

We also investigate the angular momenta of the dark matter and the hot or cold gas that are in or around a given dark matter halo. At a given snapshot, these angular momenta are measured in the centre-of-mass frame of the stellar particles within the selection aperture defined previously, and the reference point of angular momentum is the centre of mass of these particles.

\subsection{Semi-analytic calculations of angular momentum accretion}
\label{sec:method_SA_j_model}

The SA models that we consider in detail in this work assume that all of the gas accreted onto a dark matter halo that is not already in the form of cold gas in a galaxy immediately joins the hot gas halo of the main halo. 
Gas then cools from this hot halo and is accreted by the galaxy at the centre of the dark matter halo. Typically the gas has non-zero angular momentum, so there is an angular momentum flow accompanying the mass flow of cooling gas. Different SA models calculate this angular momentum flow in different ways.

Note that we are assuming in the SA models that the hot gas in satellite haloes is instantaneously stripped by ram pressure effects as soon as the satellite halo falls into the main halo. In reality, the stripping of hot gas from satellite haloes by both ram pressure and tidal effects may happen somewhat more gradually.

\subsubsection{The \MORGANA model} 
\label{sec:morgana_model}
The treatment of angular momentum in the \MORGANA model \citep{morgana1} is at the least sophisticated end of the  spectrum of SA models. It does not explicitly follow the angular momentum flow induced by gas cooling, but simply assumes that the specific angular momentum of a central galaxy is always equal to the mean specific angular momentum of its host dark matter halo, $\bar{\boldsymbol{j}}_{\rm halo}=\boldsymbol{J}_{\rm halo}/M_{\rm halo}$, where $\boldsymbol{J}_{\rm halo}$ and $M_{\rm halo}$ are respectively the total angular momentum and mass of the dark matter halo. This assumption is also adopted in some simple calculations of galaxy disk sizes \citep[e.g.][]{disk_size_mo_1998}. Many previous works that compared hydrodynamical simulations with SA models [e.g.\ \citet{Stewart_j_2013,Danovich_j_2015} and also see \S4.2.4 of \citet{Somerville_review_2015}]  focused mainly on this type of SA calculation of angular momentum.

\SAG\citep{SAG_model_2018}, \SAGE\citep{SAGE_model_2016} and \MERAXES\citep{MERAXES_model_2016} adopt treatments similar to that in the \MORGANA model. They assume an exponential disk with scale radius  $r_{\rm d}=\lambda_{\rm halo}r_{\rm vir}/\sqrt{2}$, where $r_{\rm vir}$ is the halo virial radius and $\lambda_{\rm halo}=|\bar{\boldsymbol{j}}_{\rm halo}|/(\sqrt{2}r_{\rm vir}V_{\rm vir})$ is the halo spin parameter. This equation for $r_{\rm d}$ [taken from equation~(12) of \citet{disk_size_mo_1998}, also see \citet{Fall_1980_disk_size}] is obtained by requiring that the specific angular momentum of the disk equals that of the host halo, assuming a non-self-gravitating disk in a singular isothermal halo.

All the above-mentioned models mainly focus on the magnitude of the galaxy specific angular momentum. But since they calculate this from $\bar{\boldsymbol{j}}_{\rm halo}$, in this work, we adopt the direction of $\bar{\boldsymbol{j}}_{\rm halo}$ as the direction of the specific angular momentum of the corresponding central galaxy for these models. $\bar{\boldsymbol{j}}_{\rm halo}$ is provided by the N-body merger trees extracted from the dark matter only simulation.

\subsubsection{The \lgalaxy model} 
\label{sec:lgalaxies_model}
The \lgalaxy model \citep[e.g.][]{munich_model_Guo11} calculates the angular momentum accretion by assuming that in any timestep the specific angular momentum of the cooled-down gas equals the mean value for the halo $\bar{\boldsymbol{j}}_{\rm halo}$. Note that here the  angular momentum of cooled-down gas is treated as a full vector. Then one has $\Delta\boldsymbol{J}_{\rm cool}=\Delta M_{\rm cool}\bar{\boldsymbol{j}}_{\rm halo}$, where $\Delta\boldsymbol{J}_{\rm cool}$ is the total angular momentum brought in by the cooled-down gas, and $\Delta M_{\rm cool}$ is the mass of the gas cooled down in this timestep. $\Delta\boldsymbol{J}_{\rm cool}$ is accumulated for each galaxy, and this gives the evolution of a galaxy's angular momentum.

The \DARKSAGE model in \citet{Dark_SAGE_model_2016} also assumes $\Delta\boldsymbol{J}_{\rm cool}=\Delta M_{\rm cool}\bar{\boldsymbol{j}}_{\rm halo}$. The \DARKSAGE model further assumes that the gas cooled down in a single timestep forms a disk with certain angular momentum distribution, and adds this disk to the disk of the central galaxy to calculate the evolution of the disk structure. In the present work and also in many other SA models, however, we focus on global properties such as mean specific angular momentum of the central galaxy, so we are not concerned with its detailed distribution within a galaxy. At the level we are considering, the \DARKSAGE model for angular momentum accretion is equivalent to the \lgalaxy model.

Note that a later \DARKSAGE paper \citep{Dark_SAGE_Extension_2018} adopts a different angular momentum distribution for the gas cooled down in a single timestep. With this new form, $\Delta\boldsymbol{J}_{\rm cool}=\Delta M_{\rm cool}\bar{\boldsymbol{j}}_{\rm halo}$ no longer holds. This form is based on a fit to results from the \eagle hydrodynamical simulation  \citep{Stevens_cooling_2017}. However, this fit is only for haloes possessing a hot gas halo, while here we consider angular momentum accretion in both hot gas halo and filamentary accretion regimes, so we postpone the consideration of this update, which is based on a specific accretion regime, to later works.

The SA model \SHARK\citep{SHARK_model_2018} assumes that for the gas cooled down in a single timestep, $|\Delta\boldsymbol{J}_{\rm cool}|=\Delta M_{\rm cool}|\bar{\boldsymbol{j}}_{\rm halo}|$, and treats angular momentum as a scalar, meaning that during the evolution of a halo the angular momentum does not change its direction. The assumption about the magnitude of the angular momentum  is the same as in the \lgalaxy model, but the assumptions about its direction are different in these two models. We will briefly discuss the possible effects of the assumptions about angular momentum direction in the \SHARK model in \S\ref{sec:comparison}.

\subsubsection{The \GALFORM models} 
\label{sec:galform_model}
The treatments of angular momentum in the various versions of \GALFORM are more sophisticated than those in the above two models. They first follow the evolution of the total angular momentum of the assumed spherical hot gas halo. As mentioned above, all diffuse gas (i.e. not in the form of cold gas in galaxies) newly accreted onto a dark matter halo is assumed to join the hot gas halo, and so change its total angular momentum. The \GALFORM models assume that the gas accreted onto the hot gas halo in a given timestep has the same specific angular momentum as the dark matter accreted at the same timestep, i.e.\ $\boldsymbol{j}_{\rm gas,new}=\boldsymbol{j}_{\rm dark,new}$, where $\boldsymbol{j}_{\rm gas,new}$ and $\boldsymbol{j}_{\rm dark,new}$ are the specific angular momenta of the newly accreted gas and dark matter respectively. Further, $\boldsymbol{j}_{\rm dark,new}=\Delta\boldsymbol{J}_{\rm halo}/\Delta M_{\rm halo}$, where $\Delta\boldsymbol{J}_{\rm halo}$ and $\Delta M_{\rm halo}$ are respectively the change in angular momentum and mass of the dark matter halo in that timestep. It is easy to calculate $\boldsymbol{j}_{\rm dark,new}$ from the halo merger trees in dark matter only simulations. 

Note that $\boldsymbol{j}_{\rm dark,new}$ includes the angular momentum of dark matter in satellite haloes that are accreted onto the main halo, and because we adopt the instantaneous stripping assumption in all SA models considered here, $\boldsymbol{j}_{\rm gas,new}$ also includes the contributions from the halo gas in these accreted haloes.

When some gas is removed from the hot gas halo by cooling, its associated angular momentum is calculated from the radius of the corresponding gas shell together with an assumed radial distribution of the specific angular momentum of the hot gas halo, $\boldsymbol{j}_{\rm hot}(r)$. This angular momentum is subtracted from the total angular momentum of the hot gas halo to correspond to the removal of gas. All \GALFORM cooling models assume that $\boldsymbol{j}_{\rm hot}$ at different radii has the same direction, which is the direction of the total angular momentum of the spherical hot gas halo. This direction can change when new gas is added to the hot gas halo. Different \GALFORM cooling models assume different forms for $\boldsymbol{j}_{\rm hot}(r)$, and have different treatments of the hot gas halo and the removed gas. This is described next. In all of the \GALFORM cooling models, the gas that has cooled is assumed to conserve its angular momentum during the infall from the hot gas halo to the central galaxy.

Note that in order to be compatible with both Monte Carlo and N-body merger trees, previous \GALFORM models ignored the directions of halo angular momenta provided by N-body simulations, and assumed that the galaxy and halo angular momenta remained perfectly aligned at all times. Under this assumption the angular momentum can be treated as a scalar. Since in this work we only use merger trees from N-body simulations, we remove this limitation and use the full information on the angular momentum vector provided by these simulations.

\begin{enumerate}

\item \GALFORM {\it \ model GFC1}

The GFC1 (GalForm Cooling 1) cooling model was introduced in \citet{galform_bower2006} (building on \citet{cole2000}), and has been widely used in recent \GALFORM models, \citep[e.g.][]{galform_gonzalez2014,galform_lacey2015}. A full description of it can be found in \citet{new_cool}.

This model makes use of so-called halo formation events, which are defined recursively as the time when a halo becomes at least twice as massive than its main progenitor at the last halo formation event. The hot gas halo is assumed to be nearly static between two halo formation events, and its properties are reset at each of these events. It assumes $\boldsymbol{j}_{\rm hot}(r)\propto r$, with the normalization being determined by the total angular momentum of the hot halo. This form for $\boldsymbol{j}_{\rm hot}(r)$ is motivated by some hydrodynamical simulations without radiative cooling \citep{cole2000}.

At each timestep, the gas removed from the hot gas halo is that having enough time to both radiate all of its thermal energy and to fall under gravity to the halo centre. Its mass and angular momentum are removed from the hot gas halo, and are added to the central galaxy in this halo.

\item \GALFORM {\it \ model GFC2}

The GFC2 (GalForm Cooling 2) model was introduced in \citet{benson_bower_2010_cooling}. More detailed descriptions of it are given in that paper and in \citet{new_cool}.

This model removes the artificial halo formation events, and assumes that the hot gas halo gradually contracts as cooling proceeds. The introduction of this contraction makes this model more physically realistic than the GFC1 model, because the gas cooling typically starts from the halo centre, and when gas has cooled down it would no longer provide pressure support to the gas at larger radii.

\citet{benson_bower_2010_cooling} assumed that $\boldsymbol{j}_{\rm hot}(r)\propto r$ at the start of cooling in a halo, but at later times the cooling-induced contraction changes $\boldsymbol{j}_{\rm hot}(r)$. They tried to derive the contraction induced evolution of $\boldsymbol{j}_{\rm hot}(r)$ under the assumption that each Lagrangian shell of hot gas halo conserves its angular momentum under contraction. To achieve this, \citeauthor{benson_bower_2010_cooling} adopted a specific functional form for the specific angular momentum as a function of enclosed baryonic mass (normalized to the baryonic mass within the virial radius), $\boldsymbol{j}_{\rm hot}[M(<r)]$, instead of for $\boldsymbol{j}_{\rm hot}(r)$. However, the enclosed mass $M(<r)$ labels Lagrangian shells only when the mass within each radius is constant, and this is not satisfied in the cosmological structure formation context, in which the gas newly joining the hot gas halo changes $M(<r)$. Since this more complex choice does not fully achieve its aim, here for simplicity we instead assume $\boldsymbol{j}_{\rm hot}(r)\propto r$ for all times. Note that this does not imply that this model would give the same results as the GFC1 model, because they are still different for other aspects of the cooling calculation.

As in the GFC1 model, the gas removed from the hot gas halo at each timestep is that having enough time to both radiate all of its thermal energy and to fall under gravity to the halo centre.

\item {\it New} \GALFORM {\it \ cooling model}

This model was introduced in \citet{new_cool} and a detailed description is provided there.

Like GFC2, this new model dispenses with artificial halo formation events. It also includes the contraction of the hot gas halo, but in a more physically consistent way than in the GFC2 model. This model assumes $\boldsymbol{j}_{\rm hot}(r)\propto r$ for gas before it has started cooling, including both the hot gas halo at the initial time and also the gas newly accreted onto a dark matter halo at each timestep, which is assumed to be newly heated up. This is more physically reasonable than in the GFC1 model, because the dependence $\boldsymbol{j}_{\rm hot}(r)\propto r$ is suggested by hydrodynamical simulations without radiative cooling.

This model calculates the evolution of $\boldsymbol{j}_{\rm hot}(r)$ induced by the halo contraction under the assumption that each Lagrangian hot gas shell conserves its angular momentum during the contraction. This assumption leads to an equation for the angular momentum profile after the contraction, and the  model solves this to derive the new $\boldsymbol{j}_{\rm hot}(r)$.

Once a hot gas shell has radiated away all of its thermal energy, it is removed from the hot gas halo. (This is different from what is assumed in the GFC1 and GFC1 models, in which a free-fall condition is also applied, as described above.) Its mass and angular momentum are then added to a halo cold gas reservoir, which drains onto the central galaxy on the free-fall timescale. The angular momentum drained from this reservoir is assumed to be proportional to the mass drained out, with the proportionality factor being the mean specific angular momentum of this reservoir, $\bar{\boldsymbol{j}}_{\rm halo,cold}=\boldsymbol{J}_{\rm halo,cold}/M_{\rm halo,cold}$, where $\boldsymbol{J}_{\rm halo,cold}$ and $M_{\rm halo,cold}$ are respectively the total angular momentum and mass of the halo cold gas reservoir. The drained mass and angular momentum are added to the central galaxy.

Note that if the cooling timescale is much longer than the free-fall timescale, then the mass in the halo cold gas reservoir remains small, because the draining timescale is then shorter than the feeding timescale, while in the opposite situation, the mass in the halo cold gas reservoir could be significant. For the latter case, the introduction of this reservoir leads to a more physically realistic treatment of the hot gas halo than in either of the GFC1 and GFC2 models, because now the gas that cooled down but not yet been accreted by the central galaxy is removed from the hot gas halo, rather than being left in this hot halo as in the GFC1 and GFC2 models.

\end{enumerate}

\subsubsection{Further details}
\label{sec:notes_for_SA_models}

In the hydrodynamical simulation there is a temperature threshold for cooling (equation~\ref{eq:T_lim_cool}) to restrict cooling to happen only in well-resolved haloes. Correspondingly, in all of the SA models in this work, radiative cooling is only allowed in haloes more massive than $2\times 10^{10}\Msol$.

All of the SA models considered in this work calculate the accretion of angular momentum in gas and dark matter into the halo based on the angular momentum of the dark matter haloes. The dark matter halo merger trees used in this work are generated using a dark matter only simulation that has the same initial conditions for the total density as in the hydrodynamical simulation. For each halo in which gas cooling is allowed, we directly adopt its angular momentum vector measured from the dark matter only simulation. As pointed out in \citet{Bett_2007}, the measurement of halo angular momentum in simulations is only robust when the halo is resolved with at least $300$ particles. This requirement is met here, because the cooling is restricted to haloes with $M_{\rm halo}\geq 2\times 10^{10}\Msol$, which are resolved with at least $2000$ particles.

As mentioned in \S\ref{sec:method_j_measurement}, in the hydrodynamical simulation the contribution to the cumulative mass and angular momentum of the cooled-down gas from snapshots having massive satellite mergers is removed. Correspondingly, the SA model predictions also exclude the data from the corresponding timesteps.

\section{Results} 
\label{sec:results}
\subsection{Assumptions involved in SA calculation of angular momentum accretion}
\label{sec:assumption_check}
The \lgalaxy and \GALFORM models calculate the angular momentum accretion onto central galaxies based on several assumptions. We check the validity of these, before comparing the predicted cumulative angular momentum of the cooled-down gas to the hydrodynamical simulation. Here, for the \GALFORM models, we mainly focus on the new cooling model, because it is the most physically realistic one among the three \GALFORM cooling models.

\subsubsection{Halo samples}
\label{sec:assumption_check_halo_sample}

\begin{figure*}
\centering
\includegraphics[width=1.0\textwidth]{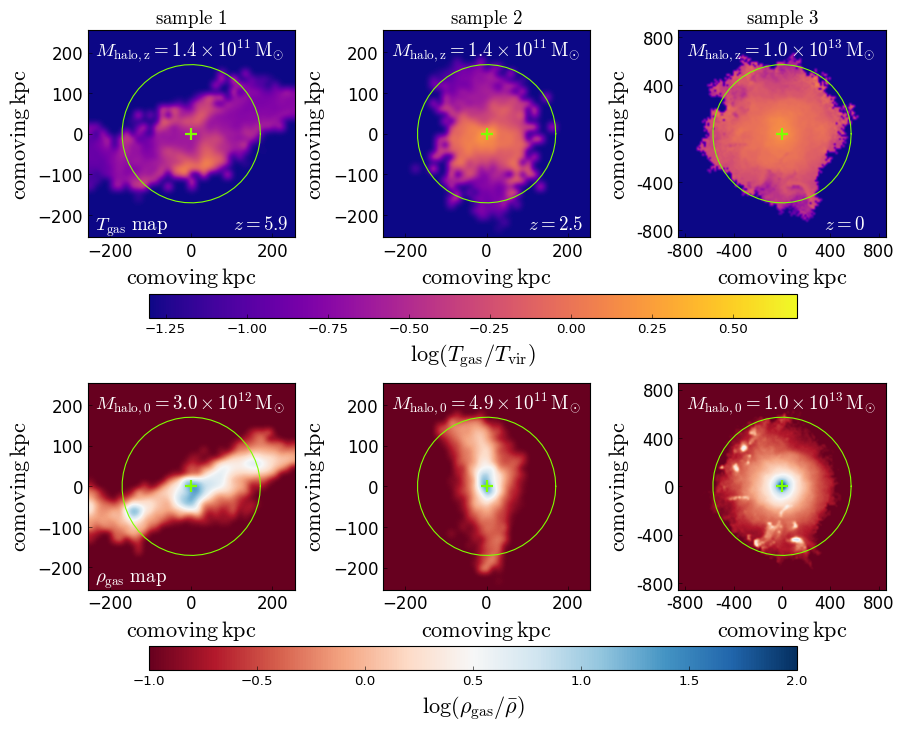}
\caption{Projected gas temperature (upper row) and density (lower row) maps for three example haloes from the three halo samples defined in \S\ref{sec:assumption_check_halo_sample}. The colour scales are given at the bottom of corresponding rows. Each column is for a given example halo, with the sample name, selection redshift, halo mass at that redshift ($M_{\rm halo,z}$) and $z=0$ descendant halo mass ($M_{\rm halo,0}$) given in the corresponding column. In each panel, the green cross indicates the halo centre, while the green circle shows the virial radius calculated using the Dhalo mass of the corresponding halo. The gas in haloes from sample 1 and sample 2 is clearly filamentary and cold ($T_{\rm gas}<T_{\rm vir}$). The halo from sample 3 shows a roughly spherical hot gas halo with temperature around $T_{\rm vir}$. }\label{fig:example_halo}

\end{figure*}

\begin{table*}
\centering

\caption{Basic information on halo samples}

\begin{tabular}{c|ccccccc}
\hline
sample name & $N_{\rm halo}^a$ & $z^b$ & $M_{\rm halo,z,low}^c$ & $M_{\rm halo,z,high}^d$ & $M_{\rm halo}(z=0)^e$ & $M_{\rm halo}(z=0)$ & $M_{\rm halo}(z=0)$ \\
&&& $[{\rm M}_{\odot}]$ & $[{\rm M}_{\odot}]$ & median $[{\rm M}_{\odot}]$ & $10$ percentile $[{\rm M}_{\odot}]$ & $90$ percentile $[{\rm M}_{\odot}]$ \\
\hline
sample 1 & $38$ & $5.9$ & $7\times 10^{10}$ & $1.5\times 10^{11}$ & $2.5\times 10^{12}$ & $1.2\times 10^{12}$ & $1.3\times 10^{13}$ \\
sample 2 & $392$ & $2.5$ & $7\times 10^{10}$ & $1.5\times 10^{11}$ & $3.7\times 10^{11}$ & $1.8\times 10^{11}$ & $8.3\times 10^{11}$ \\
sample 3 & $49$ & $0$ & $5\times 10^{12}$ & $\infty$ & $9.5\times 10^{12}$ & $6.1\times 10^{12}$ & $3.8\times 10^{13}$ \\
\hline
\end{tabular}

Notes: (a) total number of haloes in sample; (b) redshift at which  sample selected; (c) and (d) lower and upper mass bounds of haloes in sample at selection redshift; (e) $z=0$ descendant masses of haloes in sample.

\label{table:halo_sample_info}
\end{table*}

SA models typically assume that the gas accreted onto a dark matter halo first forms a hot, roughly hydrostatic gas halo and then gas cools from this and falls onto the central galaxy. In hydrodynamical simulations, however, as pointed out by many previous works \citep[e.g.][]{cold_accretion_keres,dekel_filament_accretion,Nelson_IGM_sim}, this is only true for massive, low-redshift haloes, while high-redshift, low mass haloes mainly accrete through thin filaments, and the cold, dense gas flows along filaments are hard to heat up and remain cold when they fall freely onto central galaxies. The transition from the filamentary accretion to the hot gas halo happens at a halo mass scale about $3\times 10^{11}\Msol$, and this scale is insensitive to redshift \citep{cold_accretion_keres}.

As shown in Paper~I, SA models assuming spherical hot gas haloes nonetheless predict mass accretion rates for cool gas flowing into halo centres that are similar to those given by filamentary accretion in hydrodynamical simulations. This is because for high-redshift, low-mass haloes, the cooling timescale of the assumed hot gas halo is short and the mass flow rate is still mainly determined by the gravitational infall timescale, just as in the filamentary accretion case.

However, calculations of the angular momentum flow involve additional assumptions over those made for the cooling mass flow calculation, so the calculation of the angular momentum accretion rate based on a model of a spherical hot halo needs to be checked separately. This motivates us to create separate halo samples for the filamentary accretion regime and the hot gas halo regime.

More specifically, we built three halo samples. In {\it sample~1}, we selected haloes at $z=5.9$ with $7\times 10^{10}\Msol\leq M_{\rm halo} \leq 1.5\times 10^{11}\Msol$, and in {\it sample~2} we selected haloes at $z=2.5$ in the same mass range. Haloes in these two samples are in the filamentary accretion regime, but they have very different $z=0$ descendant halo masses, $M_{\rm halo}(z=0)$, and so sample different environments. The $z=0$ descendants in sample~1 cover a 10-90\% mass range $[1.2\times 10^{12}\Msol,\ 1.3\times 10^{13}\Msol]$, with median $2.5\times 10^{12}\Msol$, while for sample~2 the range and median are $[1.8\times 10^{11}\Msol,\ 8.3\times 10^{11}\Msol]$ and $3.7\times 10^{11}\Msol$ respectively. This indicates that the haloes in sample~1 are in relatively high density regions, and haloes in sample~2 are in low density regions. In {\it sample~3}, we selected haloes at $z=0$ with $M_{\rm halo}\geq 5\times 10^{12}\Msol$. These haloes are in the hot gas halo regime. 

Note that in all three samples, we excluded haloes that have had major mergers (mass ratio larger than $1:2$) within previous four snapshots (corresponding roughly to one halo dynamical timescale), in order to remove strongly disturbed and unrelaxed systems. The removed haloes are only a minor fraction of the corresponding sample.

Some basic information on the three halo samples is summarized in Table~\ref{table:halo_sample_info}, and three example haloes for these samples are shown in Fig.~\ref{fig:example_halo}.

\subsubsection{Filamentary accretion regime}
\label{sec:assumption_check_filament_regime}
\noindent{\bf Specific angular momentum of material about to join a halo:}
Both the new cooling model and the GFC1 and GFC2 models track the angular momentum of the gas falling into a dark matter halo. They do this by assuming that the gas and dark matter accreted onto a halo have the same specific angular momentum. We first check this assumption in the filamentary accretion regime.

We take a thin spherical shell centred on the centre of a given halo (the centre given by \SUBFIND of the most massive subgroup in that halo) and ranging from $r_{\rm vir}$ to $1.1r_{\rm vir}$, where $r_{\rm vir}$ is the virial radius of the dark matter halo. Then we select all gas cells and dark matter particles within this shell and having $\boldsymbol{V}\cdot \boldsymbol{r}<0$, where $\boldsymbol{V}$ is the velocity of a given cell or particle in the centre-of-mass frame of the stellar particle group defined in \S\ref{sec:method_j_measurement}\footnote{We checked that this frame usually is very close to the centre-of-mass frame of the whole halo.} and $\boldsymbol{r}$ is the position vector of this cell or particle relative to the halo centre. We treat these selected gas cells and dark matter particles as the gas and dark matter that are going to be accreted by the halo.

\begin{figure*}
\centering
\includegraphics[width=1.0\textwidth]{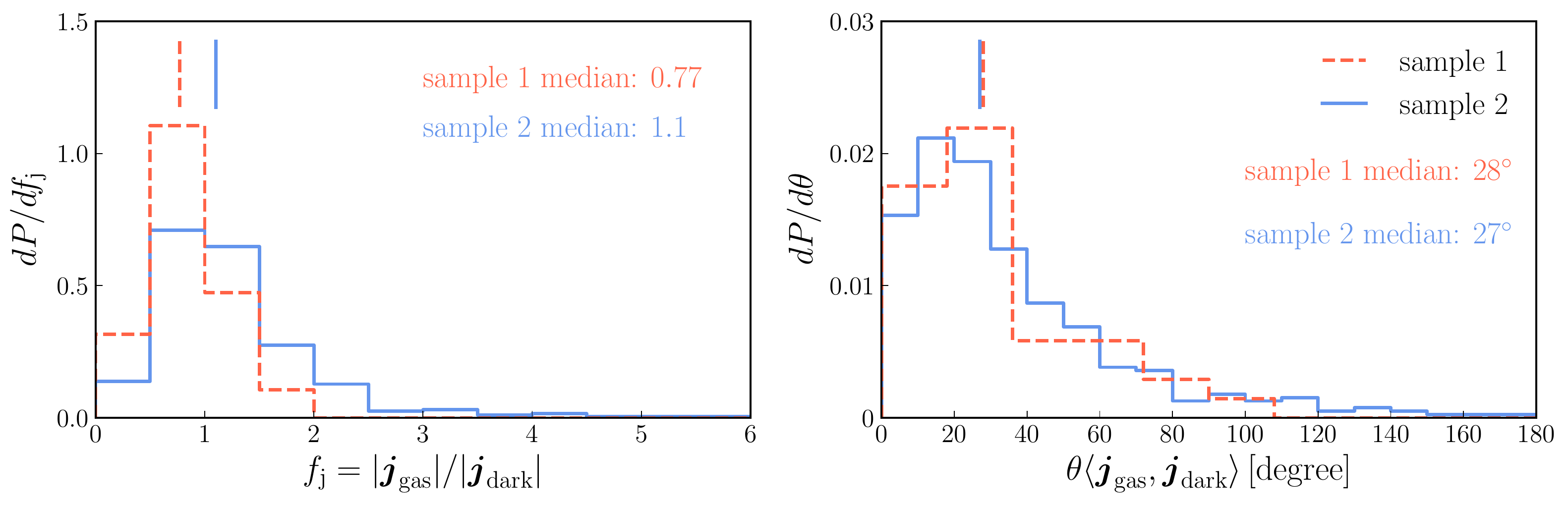}
\includegraphics[width=1.0\textwidth]{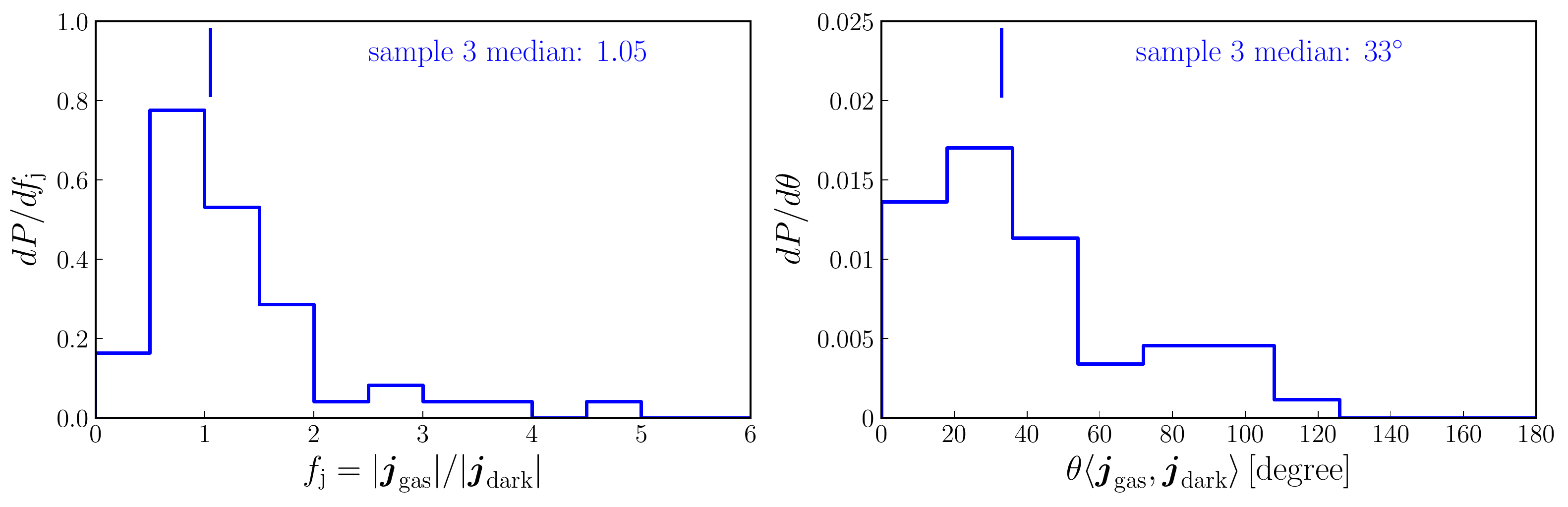}
\caption{Comparison of the specific angular momenta of the gas and dark matter that are about to be accreted onto a dark matter halo for the halo samples defined in Table~\ref{table:halo_sample_info} in \S\ref{sec:assumption_check_halo_sample}. The specific angular momenta are calculated for all inflowing gas and dark matter in the shell $r_{\rm vir} < r < 1.1r_{\rm vir}$. The upper row is for haloes in the filamentary accretion regime, while the lower row for haloes in the hot gas halo regime. The left column shows the probability distributions of angular momentum magnitude ratio, while the right column shows the probability distributions of the angle between these two angular momenta. The medians of the magnitude ratio and the direction offset for each halo sample are given in the corresponding panel, and also indicated graphically as short vertical  lines. In the upper row, different line styles indicate different halo samples, with the corresponding sample names given in the key in the right panel.}\label{fig:j_new_acc}
\end{figure*}

The upper row of Fig.~\ref{fig:j_new_acc} compares the mean specific angular momenta of the selected gas cells and dark matter particles of the two halo samples in the filamentary accretion regime. If these two angular momenta were exactly the same, then their magnitude ratio would be $1$ and angular offset $0^{\circ}$. In fact, as shown in Fig.~\ref{fig:j_new_acc}, both the magnitude ratio and angular offset have distributions. The median ratio of magnitudes is $0.77$ for sample 1 and $1.1$ for sample 2, while the median angular offset is $28^{\circ}$ for sample 1 and $27^{\circ}$ for sample 2. These medians therefore do not strongly deviate from the condition for $\boldsymbol{j}_{\rm gas}=\boldsymbol{j}_{\rm dark}$, and this is consistent with the conclusion in \citet{Danovich_j_2015}, who performed a similar comparison in the filamentary accretion regime. 

However, medians alone are not enough to prove that the assumption $\boldsymbol{j}_{\rm gas}=\boldsymbol{j}_{\rm dark}$ is valid for the SA calculation of angular momentum accretion in individual haloes, because the distributions in Fig.~\ref{fig:j_new_acc} are not particularly narrow, and the scatter around medians should also play a role in SA calculations. For a given halo the deviations from the median relation could be either correlated or uncorrelated in time. In the first case, the angular momentum accretion in some haloes would be constantly biased to one side of the median, and for the median for all haloes to remain close to $\boldsymbol{j}_{\rm gas}=\boldsymbol{j}_{\rm dark}$, there should be an approximately equal number of haloes in which the angular momentum accretion is constantly biased in the opposite sense relative to the median. When applied to a sample of haloes, the scatter shown in Fig.~\ref{fig:j_new_acc} may then lead to biases in SA predictions for properties of individual galaxies, but for the galaxy population corresponding to the halo sample, the medians of their properties are unlikely to be strongly affected. In the second case, when following the evolution histories of individual haloes, the deviations should fluctuate around the median, which approximately corresponds to $\boldsymbol{j}_{\rm gas}=\boldsymbol{j}_{\rm dark}$. In this case, when integrated in time to derive cumulative quantities, such as the total cooled-down angular momentum, the deviations for individual haloes tend to partially cancel out, and so, comparing to the first case discussed above, the SA predictions of properties for individual galaxies should be more accurate, and the median properties of a galaxy population should be less affected by these deviations. 

Applications of SA models generally involve predictions for statistical properties of galaxy populations, and because the scatter around $\boldsymbol{j}_{\rm gas}=\boldsymbol{j}_{\rm dark}$ should not strongly affect the medians derived from galaxy populations, such predictions should generally remain valid, provided that feedback effects do not introduce any new deviations. However, the scatter around $\boldsymbol{j}_{\rm gas}=\boldsymbol{j}_{\rm dark}$ will contribute to the scatter in galaxy properties, so the predictions for this scatter from 
SA models that assume $\boldsymbol{j}_{\rm gas}=\boldsymbol{j}_{\rm dark}$ are less reliable than predictions for the medians.

The physical reasons for the deviations around $\boldsymbol{j}_{\rm gas}=\boldsymbol{j}_{\rm dark}$ and possible modelling to further include them in SA models are interesting topics, but they are beyond the scope of the current work, and are left for future study.

\noindent{\bf Angular momentum conservation from $r_{\rm vir}$ to $0.1r_{\rm vir}$:}
Next we consider that the new cooling model assumes that each Lagrangian shell of the assumed hot gas halo conserves its angular momentum when the hot halo contracts, and that when the cooled-down gas accretes onto the central galaxy from the cold halo gas reservoir, there is no angular momentum loss either. In summary, the new cooling model assumes that the gas conserves its angular momentum from when it crosses inside $r_{\rm vir}$ until it reaches the central galaxy. 

In principle, checking whether this is consistent with hydrodynamical simulations requires tracing the motion of each Lagrangian gas cell. This is not straightforward in a grid-based hydrodynamical simulation like the one used here. However, in the filamentary accretion regime, the gas motion is mainly in free-fall \citep[e.g.][]{cold_accretion_keres,Stewart_j_2013,Danovich_j_2015}, and this allows a simpler way of tracing: by assuming that a gas shell moves from $r=r_{\rm vir}$ to $r\sim 0$ in a time $\Delta t\sim t_{\rm dyn}=r_{\rm vir}/V_{\rm vir}$, we can roughly derive the previous position of this shell based on its current position. Note that a more accurate tracing of gas trajectories can be achieved in \AREPO by using tracer particles \citep[e.g.][]{arepo_trace_particle}, but this is both complex to implement and to analyse, so for simplicity we postpone the use of tracer particles to a future work.

More specifically, for a given halo, we first take three spherical shells with ranges $[0.1r_{\rm vir}, 0.25r_{\rm vir}]$, $[0.25r_{\rm vir}, 0.5r_{\rm vir}]$ and $[0.5r_{\rm vir}, 0.75r_{\rm vir}]$ respectively, select all gas cells within each of these shells and calculate the mean specific angular momentum (total angular momentum divided by total gas mass). Note that here we exclude the very central region with $r<0.1r_{\rm vir}$, because, according to \citet{Danovich_j_2015}, in this region the gravitational interaction with the central galaxy introduces additional complexities. (We go back to discuss the gas behaviour within $0.1r_{\rm vir}$ later in this section.) Then we move to the previous snapshot and find the progenitor of that halo, and calculate its virial radius ($r_{\rm vir,prog}$), virial velocity ($V_{\rm vir,prog}$), and halo dynamical timescale $t_{\rm dyn,prog}=r_{\rm vir,prog}/V_{\rm vir,prog}$. Based on these quantities, we calculate a radius shift $\Delta r=(\Delta t / t_{\rm dyn,prog}) r_{\rm vir,prog}$, where $\Delta t$ is the time interval between the snapshots. In our simulation, $\Delta t \sim 0.25t_{\rm dyn,prog}$, so $\Delta r \sim 0.25r_{\rm vir}$. Then we can match each of the three spherical shells in the given halo to a corresponding shells in the progenitor halo. 
We treat these three shells in the progenitor halo as containing the same gas but at an earlier time. We then calculate the specific angular momentum of gas within these shells. Comparing the specific angular momenta from the two sets of shells we can then estimate whether the angular momentum is conserved as gas flows in through the halo.

\begin{figure*}
\centering
\includegraphics[width=0.7\textwidth]{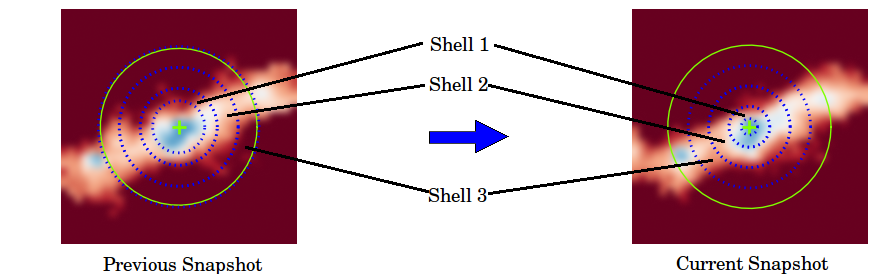}
\includegraphics[width=0.9\textwidth]{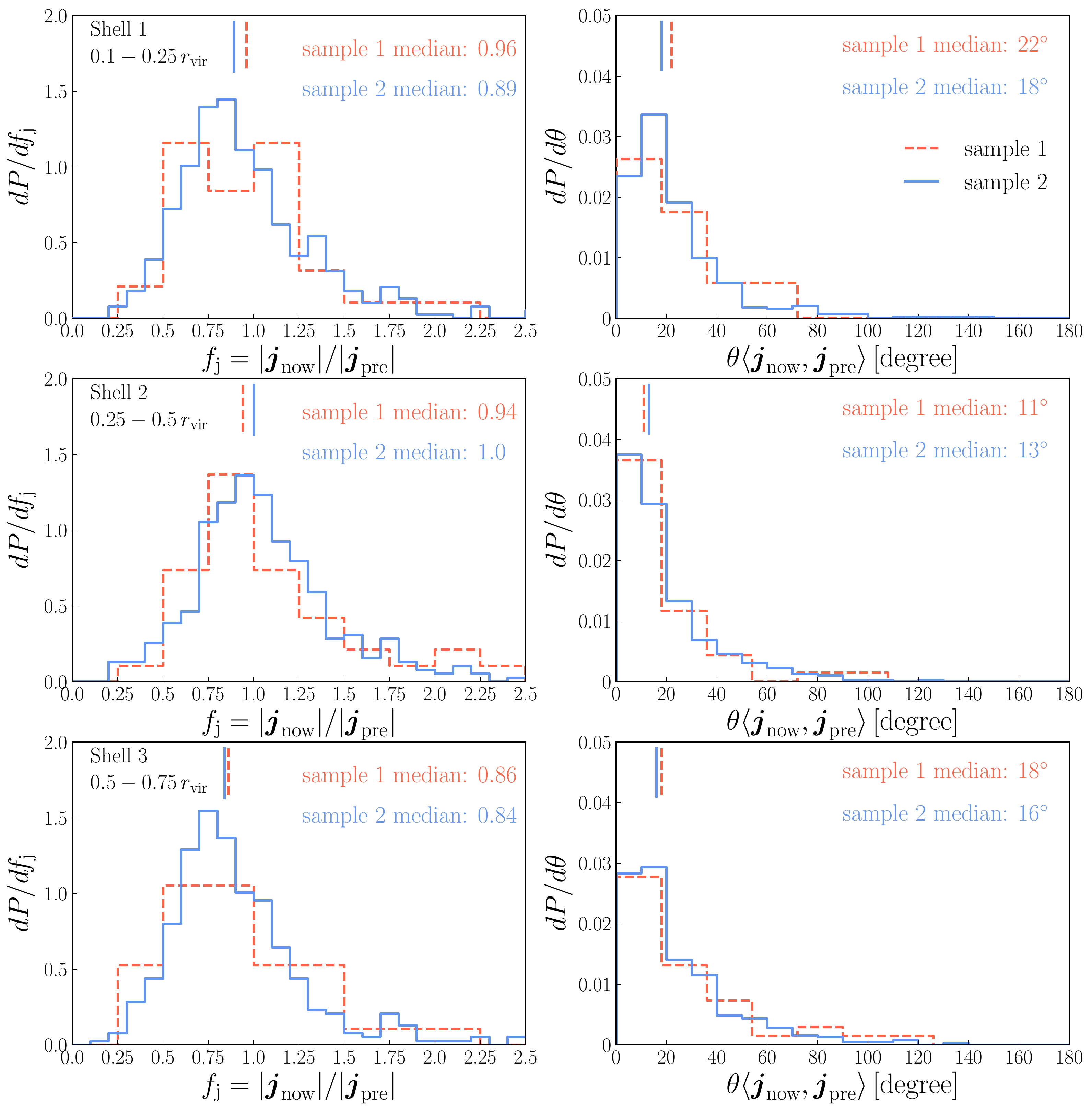}
\caption{{\it Top panels}: Depiction of how the spherical shells are defined and mapped to the previous snapshot, for haloes in the filamentary accretion regime. These shells effectively select small segments of the filamentary gas. The mapping corresponds to assuming roughly free-fall motion between snapshots, and through this the Lagrangian motion of the selected segments is approximately traced. The blue dotted circles label the shells, while the green circles indicate the virial radii of haloes, and the green crosses show the halo centres.  {\it Lower panels}: Comparison between the specific angular momenta of the gas in a shell at a given snapshot (with radii as given by the labels in each panel) and the corresponding shell at the previous snapshot. 
The gas included is that in cells belonging to the most massive subhalo in the main halo. 
The left column shows probability distributions of the angular momentum magnitude ratio, while the right column shows the probability distributions of the offset in the angular momentum direction. In each panel, different line styles are for different halo samples defined in Table~\ref{table:halo_sample_info} in \S\ref{sec:assumption_check_halo_sample}, and the corresponding sample names are given in the key in the right panel of the first row. The median values for each probability distribution are given in the corresponding panel, and also indicated graphically as short vertical lines.}
\label{fig:j_t_delay_fast_cool}
\end{figure*}

The above procedure is depicted in the top panels of Fig.~\ref{fig:j_t_delay_fast_cool}. Note that although we take spherical shells, the derived specific angular momentum is for the highly anisotropic filamentary gas, because as indicated by this figure and Fig.~\ref{fig:example_halo}, the filamentary gas is the dominant component of the halo gas.

The lower panels of Fig.~\ref{fig:j_t_delay_fast_cool} show the comparison between the specific angular momenta derived from the two sets of shells. It is clear that for both sample 1 and sample 2, the specific angular momentum of a shell and its corresponding shell $\Delta t$ earlier are approximately the same. It takes about $t_{\rm dyn}$ for a Lagrangian gas element to move from $r_{\rm vir}$ to the very central region with $r<0.1r_{\rm vir}$, and this is about $3-4\Delta t$. During this time, the element would pass in sequence the spatial regions covered by successive the spherical shells. As a rough estimate of the total effect over the time to fall from $r_{\rm vir}$ to $0.1 r_{\rm vir}$, we take the product of the median angular momentum magnitude ratios derived from each shell and add up the median angular offset of these shells. For sample 1, this gives $0.96\times 0.94\times 0.86=0.77$ and $22^{\circ}+11^{\circ}+18^{\circ}=51^{\circ}$, while for sample 2 this gives $0.89\times 1.0\times 0.84=0.75$ and $18^{\circ}+13^{\circ}+16^{\circ}=47^{\circ}$. Note that the direct addition of these angles provides a conservative upper limit for the direction offset, and the actual angular offset could be smaller.

These numbers suggest that in the rapid cooling regime, the gas may lose up to 20--30\% of its angular momentum as it cools and flows in from $r=r_{\rm vir}$ to $r< 0.1 r_{\rm vir}$. However, in view of the various approximations made in our estimate, it should not be seen as definitive. A more accurate estimate will require using tracer particles for the gas, which are also required to study this effect in the slow cooling regime. Given the complex environment in which gas accretion happens, one would not expect exact conservation of angular momentum as the gas flows in. For now, we judge that assuming angular momentum conservation during inflow is still a reasonable approximation to make in SA models, but future studies might require revisions to this simple scheme.

Our inference that there is typically little change in the angular momentum of the gas as it inflows from $r_{\rm vir}$ to $r\sim 0.1r_{\rm vir}$ is supported by the work of \citet{Danovich_j_2015} on gas accretion in the filamentary regime.
\citeauthor{Danovich_j_2015} reach this conclusion not through direct tracing of gas trajectories, but by analysing the torques on halo gas at a given snapshot. Within $\sim 0.1r_{\rm vir}$, \citeauthor{Danovich_j_2015} observed strong angular momentum loss due to the gravitational torque exerted by the central galaxy. This torque would add some of the inflowing gas' angular momentum to that of the central galaxy before it reaches the galaxy, and thus create a time difference between angular momentum accretion and mass accretion. However, the gas within $\sim 0.1r_{\rm vir}$ will be accreted by the central galaxy very quickly, so ignoring this time difference is not a serious approximation.

\begin{figure*}
\centering
\includegraphics[width=1.0\textwidth]{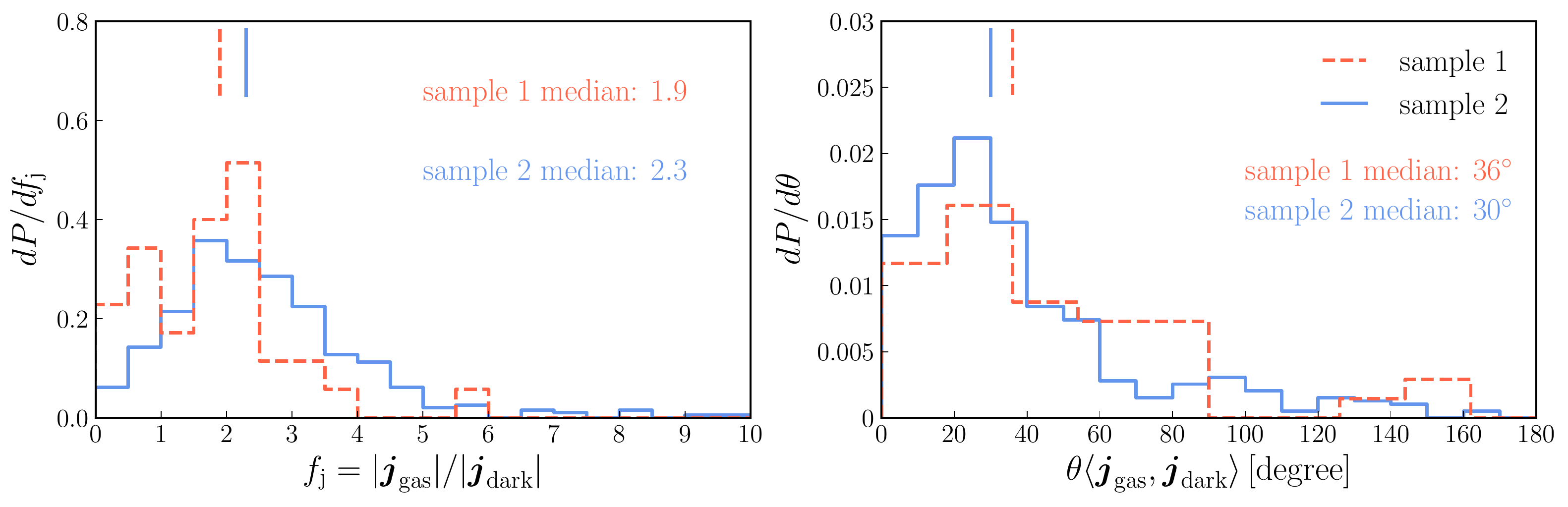}
\includegraphics[width=1.0\textwidth]{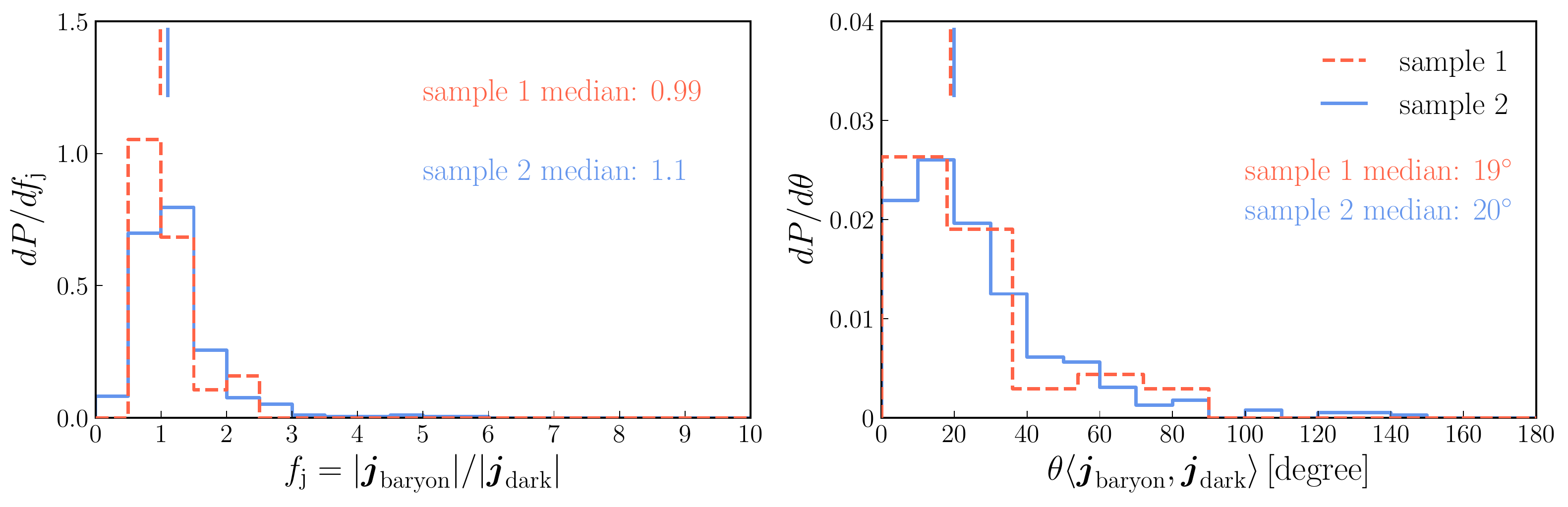}
\includegraphics[width=1.0\textwidth]{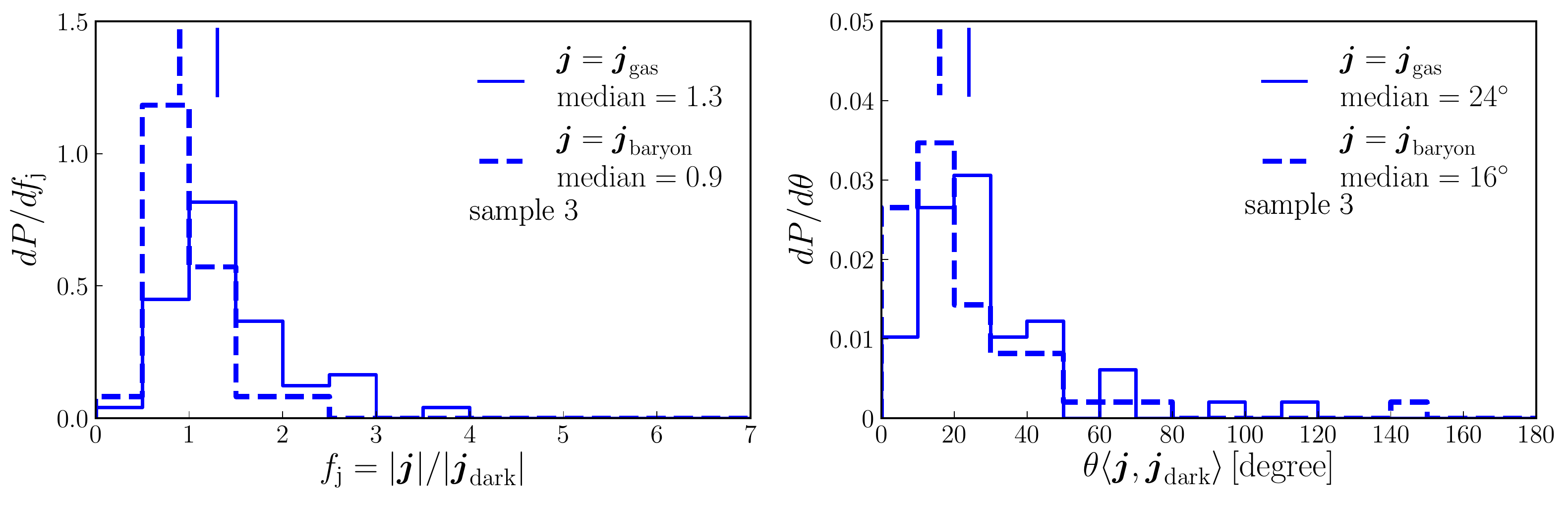}
\caption{{\it Upper and middle panels}: Comparisons between the mean specific angular momentum of the dark matter halo and of the gas within $r_{\rm vir}$ (upper row), and between that of the dark matter halo and of all baryons within the halo (middle row), for the two halo samples (defined in Table~\ref{table:halo_sample_info} in \S\ref{sec:assumption_check_halo_sample}) in the filamentary accretion regime. The dark matter, gas and stars included are those belonging to the most massive subhalo in the main halo. The left panels show the probability distributions of the angular momentum magnitude ratio, while the right panels show the probability distributions of the angular momentum direction offset. Different line styles are for different halo samples, with the corresponding sample names given in the key in the right panel. The medians of the magnitude ratio and the direction offset of each halo sample are given in the corresponding panel, and are also shown graphically as vertical short lines. {\it Lower panels}: the corresponding comparison for haloes in the hot gas halo regime (halo sample 3). The median of each distribution is given in the corresponding key and shown as a short vertical line. }
\label{fig:j_whole_halo_fast_cool}
\end{figure*}

\noindent{\bf Comparison of angular momenta of halo gas and dark matter}
We start by comparing the mean specific angular momentum of the halo gas with that of the dark matter, as has also been done in some previous works \citep[e.g.][]{Stewart_j_2013,Danovich_j_2015}. The upper row of Fig.~\ref{fig:j_whole_halo_fast_cool} compares the mean specific angular momentum of the gas within $r_{\rm vir}$ with that of the entire dark matter halo. Note that in the simulation this selection of gas would include the cold gas already in the central galaxy, but this contamination should be minor, because as described in \S\ref{sec:method_j_measurement} and Paper~I, the star formation recipe used in the hydrodynamical simulation rapidly turns most of the gas in galaxies into stellar particles. 

Fig.~\ref{fig:j_whole_halo_fast_cool} shows that the halo gas typically has a specific angular momentum around twice that of the dark matter halo. This is consistent with the results in \citet{Stewart_j_2013} and \citet{Danovich_j_2015}. The specific angular momentum of halo gas is higher than that of the dark matter halo, because recently accreted matter tends to have higher specific angular momentum than matter accreted earlier on, and while the dark matter halo is a mixture of both recently and early accreted matter, the halo gas only includes recently accreted material, with earlier accreted gas having already fallen into the central galaxy and possibly been converted into stars. Once we include the baryons in central galaxies, the mean specific angular momentum of the baryons in a halo becomes comparable to that of the dark matter in the halo, as shown in the lower row of Fig.~\ref{fig:j_whole_halo_fast_cool}.

At first glance, the results in Fig.~\ref{fig:j_whole_halo_fast_cool} seem to be contrary to the assumption made in the \lgalaxy model, in which the specific angular momentum of the gas cooled down and accreted onto a central galaxy in one timestep is set equal to the mean specific angular momentum of the dark matter halo hosting the galaxy at that time. However, one should bear in mind that the halo gas needs time to fall onto the central galaxy, and that not all the halo gas we see at a snapshot would be accreted onto the central galaxy in one timestep.  Therefore, to compare more accurately with what is assumed in SA models, we should instead use the mean specific angular momentum of the gas that would be actually accreted within a timestep. 

The timestep for SA models is the interval between two adjacent snapshots, which in our simulation is about $0.25$ halo dynamical timescales, therefore we should use the mean specific angular momentum of the gas within $0.25r_{\rm vir}$. Again, because gas in galaxies is quickly turned into stars, the contamination of this gas to the gas accreted within a timestep should be minor.

\begin{figure*}
\centering
\includegraphics[width=1.0\textwidth]{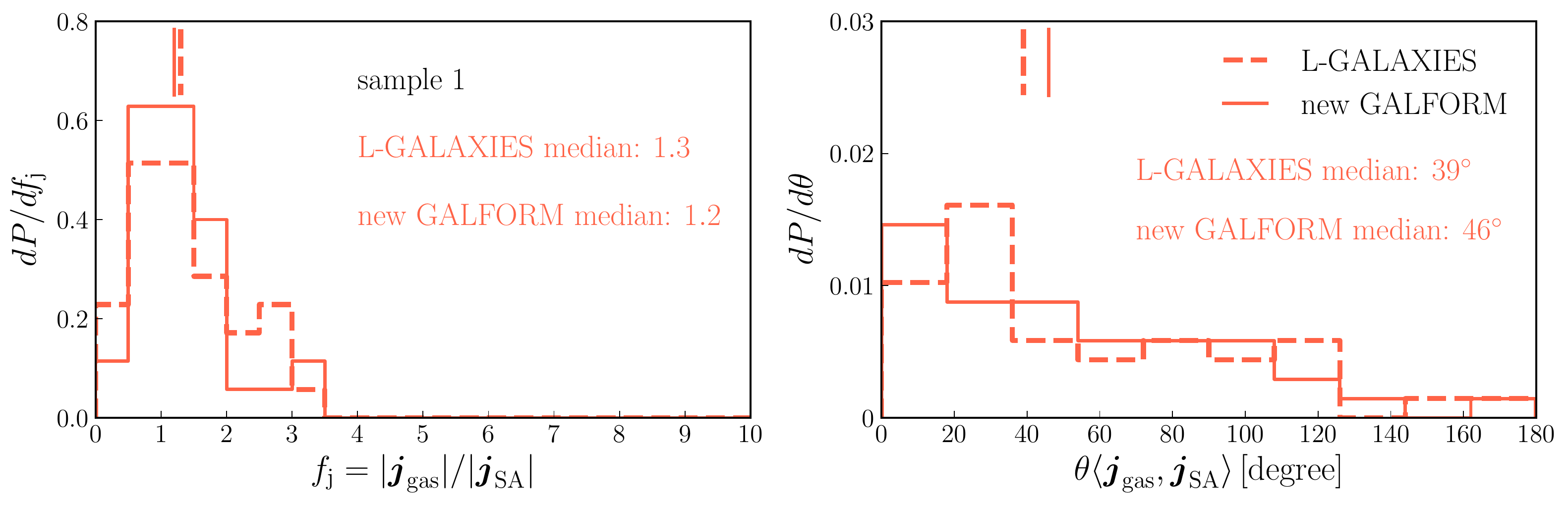}
\includegraphics[width=1.0\textwidth]{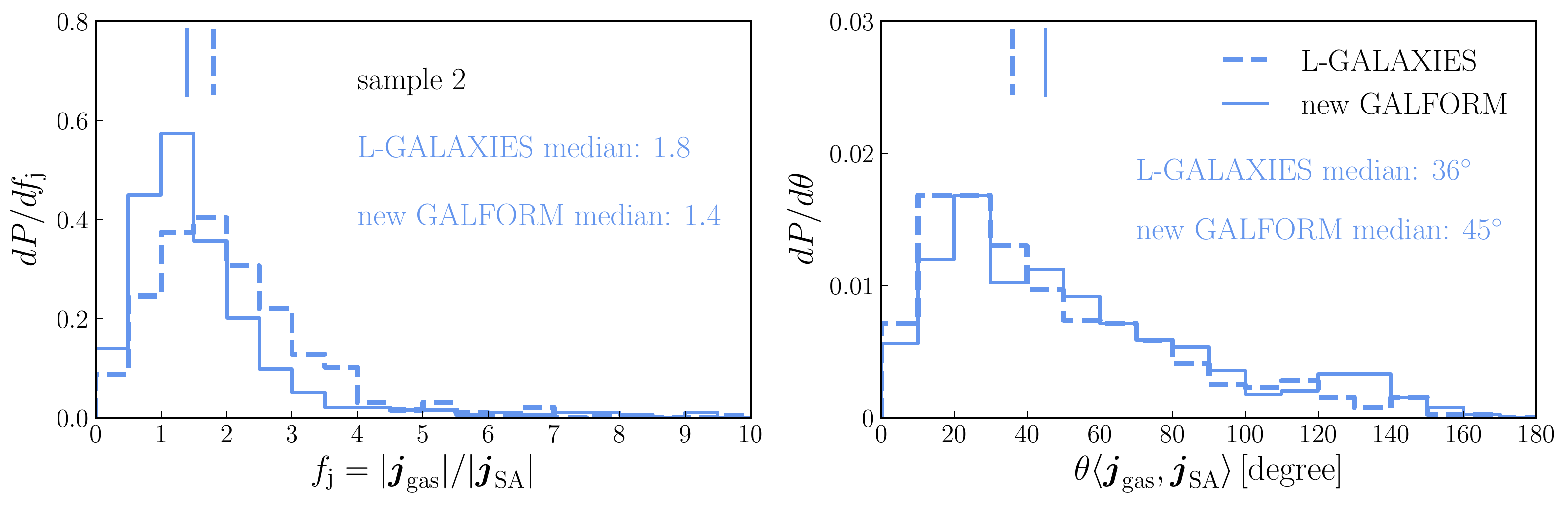}
\caption{Comparing the specific angular momentum of the gas accreted onto the central galaxy in a single timestep as calculated in the \lgalaxy and new \GALFORM cooling models with the mean specific angular momentum of the gas in the simulation within $0.25r_{\rm vir}$. For the simulation, the gas included is that belonging to the most massive subhalo in the main halo. The left panels are the probability distributions of the angular momentum magnitude ratio, while the right panels are the probability distributions of the angular momentum direction offset. The top row is for sample 1 and the bottom row for sample 2 (samples are defined in Table~\ref{table:halo_sample_info} in \S\ref{sec:assumption_check_halo_sample}). In each panel, different line styles are for different SA models, with the corresponding model names given in the key in the right panels. In each panel, the median value for each probability distribution is shown as a short vertical line, and its numerical value is also given.}
\label{fig:j_central_fast_cool}
\end{figure*}

Fig.~\ref{fig:j_central_fast_cool} compares the mean specific angular momentum of the gas in the simulation with $r < 0.25r_{\rm vir}$ with the specific angular momentum of the gas accreted onto the central galaxy in a single timestep as calculated in the \lgalaxy model and the new \GALFORM cooling model. In the \lgalaxy model, this specific angular momentum is assumed to equal that of the dark matter halo, while in the new cooling model, it is assumed to equal that of the halo cold gas reservoir. As shown in Fig.~\ref{fig:j_central_fast_cool}, at the level of medians, both of these assumptions give results that do not strongly deviate from the simulation results (with median magnitude ratio around $1.5$ and median angular offset around $40^{\circ}$), although the \lgalaxy predictions for the magnitude of the accreted angular momentum are in slightly poorer agreement with the simulation than for the \GALFORM model, particularly for sample~2, while the \GALFORM model predicts angular momentum directions that are in slightly poorer with the simulation than in the \lgalaxy model.  Fig.~\ref{fig:j_central_fast_cool} also shows considerable scatter around the medians. This scatter indicates that these assumptions may not work well for some individual haloes, but, as discussed earlier in this sub-section, this scatter is unlikely to strongly affect median properties of galaxy populations. This supports the validity of these assumptions in SA calculations of such medians, but note that these assumptions may not be as valid for calculations of the scatter in galaxy properties around the medians.

\subsubsection{Hot gas halo regime}
\label{sec:assumption_check_hot_halo_regime}
In this regime, we again first check whether the gas accreted onto a dark matter halo has the same specific angular momentum as that of the dark matter accreted onto this halo. The method of checking is the same as in \S\ref{sec:assumption_check_filament_regime}, that is, we compare the specific angular momenta of the gas and dark matter within a thin spherical shell covering $r_{\rm vir} < r < 1.1r_{\rm vir}$, and having velocities pointing towards the halo centre (i.e. falling in rather than moving out).

The bottom row of Fig.~\ref{fig:j_new_acc} shows the results of this comparison. As in the filamentary accretion regime, the specific angular momenta of infalling gas and dark matter are similar in both magnitude and direction, so the assumption made in the \GALFORM models (GFC1, GFC2 and new cooling model) is valid in both filamentary accretion and hot gas halo regimes, at least at the level of medians. The similarity of these two specific angular momenta also implies that the dark matter and total baryonic matter in a halo have similar mean specific angular momenta. However, because part of the previously accreted baryons have condensed into the central galaxy, and this part tends to have lower specific angular momentum than the baryons as a whole, the mean specific angular momentum of the hot gas halo tends to be higher than that of the dark matter. These features are shown explicitly in the bottom row of Fig.~\ref{fig:j_whole_halo_fast_cool}. 

The new \GALFORM cooling model assumes that the gas accreted onto a dark matter halo mixes homogeneously with the existing hot gas halo, and for a hot gas shell with radius $r$, the accreted gas mixed with it has specific angular momentum $\boldsymbol{j}\propto r$. It further assumes that during the contraction induced by cooling, each Lagrangian shell conserves its angular momentum. Without a tracer for Lagrangian gas elements in our current simulation, we cannot check these assumptions individually. However, we can still test the combined effect of these assumptions by comparing the radial distribution of the specific angular momentum of the hot gas, $\boldsymbol{j}_{\rm hot}(r)$, derived under these assumptions with that directly measured from the simulation.

\begin{figure*}
\centering
\includegraphics[width=0.5\textwidth]{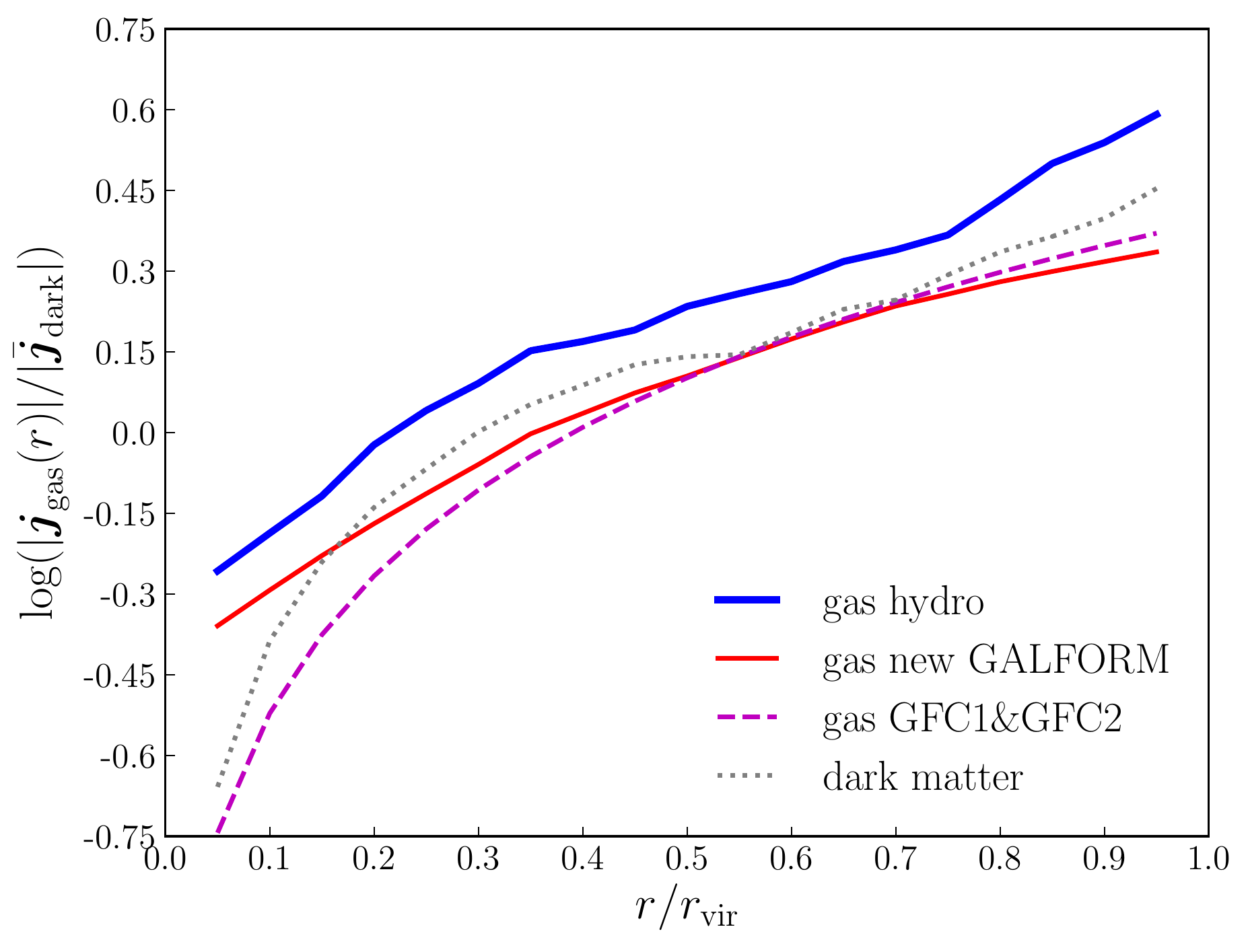}
\includegraphics[width=1.0\textwidth]{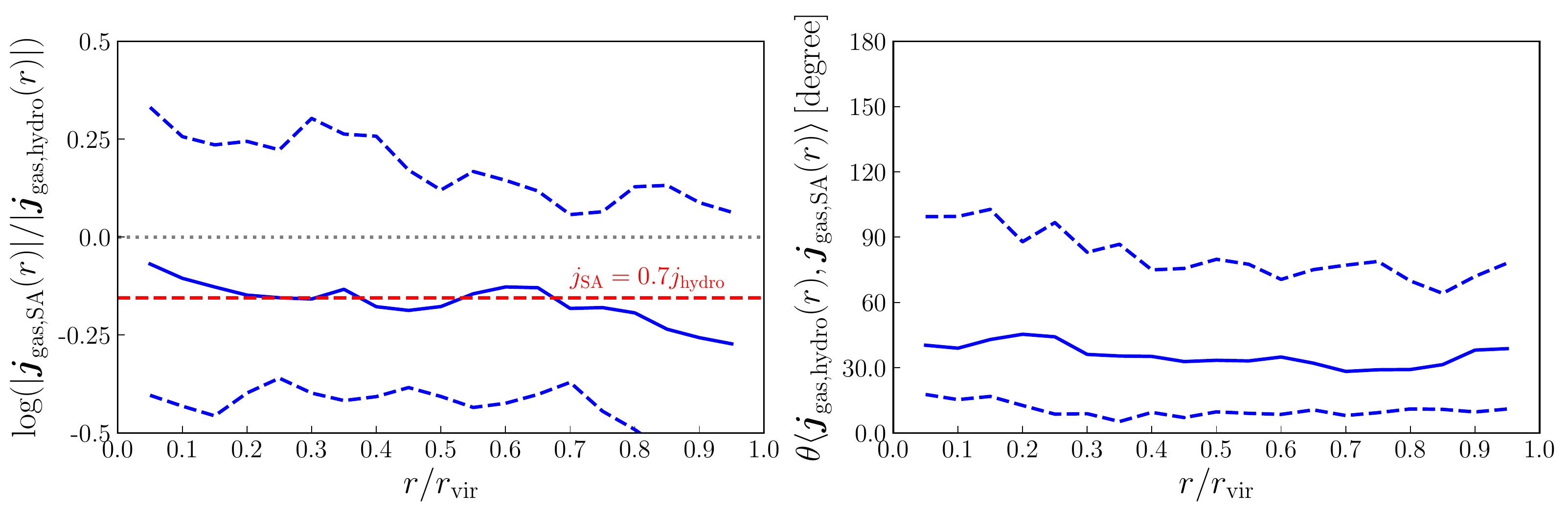}
\caption{{\it Top panel}: the median radial distributions of the magnitude of the specific angular momenta of hot gas haloes. The thick solid line shows the result measured from the simulation, the thin solid line shows the result calculated using the new \GALFORM cooling model, while the dashed line indicates the distribution for the same total angular momentum but having $|\boldsymbol{j}_{\rm hot}(r)|\propto r$, 
this being the form assumed in the older \GALFORM models GFC1 and GFC2. The dotted line shows the median distribution of specific angular momentum for the dark matter measured from the hydrodynamical simulation, and is for reference. All distributions are normalized to the mean specific angular momentum of the entire dark matter Dhalo, $\bar{\boldsymbol{j}}_{\rm dark}=\boldsymbol{J}_{\rm halo}/M_{\rm halo}$. 
{\it Lower panels}: halo by halo comparison between the specific angular momentum distributions predicted by the new \GALFORM model and that measured from the simulation. The left panel is for the magnitude ratio, and the right panel is for the direction offset. In each panel the solid line shows the median, while the dashed lines indicate the 10-90 percentile range. Note that measurements from the simulation are for gas and dark matter belonging to the most massive subhalo of each Dhalo.
}\label{fig:slow_cool_j_distri}
\end{figure*}

The top panel of Fig.~\ref{fig:slow_cool_j_distri} shows the median radial distributions of specific angular momentum of the hot gas measured from the simulation and calculated using the new cooling model. The SA model prediction  has a similar shape  to the simulation result, but with a normalization that is lower by $\sim 0.2\,{\rm dex}$. The lower two panels of Fig.~\ref{fig:slow_cool_j_distri} show the results of a halo by halo comparison. They show that the median offset in the magnitude of the specific angular momentum is about $0.1-0.2\,{\rm dex}$, 
while the median offset in direction is about $40^{\circ}$. These median offsets are only weakly dependent on $r/r_{\rm vir}$, but there is also a large halo-to-halo scatter around these medians. 

We have checked that the median magnitude offset is mainly caused by the fact that the gas density in the range $0.7r_{\rm vir}\leq r\leq r_{\rm vir}$ declines more slowly in the SA model than in the simulation. Given the shape of $|\boldsymbol{j}_{\rm hot}(r)|$, this means that hot haloes in the SA model contain a higher fraction of high specific angular momentum gas than those in the simulation. On the other hand, we found that the total angular momenta of hot haloes predicted by the SA model and the simulation are very close, despite the fact that these two methods give different predictions for cooled-down angular momenta, because in these massive haloes only a small part of the gas cools from hot haloes. Therefore, a higher fraction of high specific angular momentum gas leads to a smaller amplitude of $|\boldsymbol{j}_{\rm hot}(r)|$ in order to give the same total angular momentum of the hot gas halo. This offset between the SA model and the simulation would be reduced if a gas density profile closer to that in the simulation were used, and this is a possible future improvement of the SA model.

Considering that the SA model is highly simplified, the medians it predicts are reasonably close to those from the simulation. This supports the use of the new \GALFORM cooling model in calculating statistical properties of galaxy populations, at least at the level of median properties.

The top panel in Fig.~\ref{fig:slow_cool_j_distri} also shows the median angular momentum distribution of the hot gas calculated according to the assumptions made in the GFC1 and GFC2 models (dashed line in top panel). This distribution is obtained by assuming the same total angular momentum for the hot gas halo and the same hot gas density profile, but assuming instead $\boldsymbol{j}_{\rm hot}(r)\propto r$. 
It can be seen that the angular momentum profiles in the old and new \GALFORM cooling models are very similar in the outer parts of haloes, but the angular momentum in the old model is lower at small radii. For a halo in the simulation to possess a quasi-hydrostatic hot gas halo, its gas cooling timescale should be longer than the halo dynamical timescale, $t_{\rm dyn}$, and the accretion of cooled-down gas is then controlled by the former timescale. The time interval between two snapshots is about $0.25t_{\rm dyn}$, so if the accretion were determined by the free-fall timescale (roughly equal to $t_{\rm dyn}$), only the gas within about $0.25r_{\rm vir}$ would be accreted between snapshots. In fact, the accretion is determined by the even longer cooling timescale, so the gas accreted between snapshots should come from the halo central region, with $r<0.25r_{\rm vir}$. In this region, as shown in Fig.~\ref{fig:slow_cool_j_distri}, the $\boldsymbol{j}_{\rm hot}(r)$ assumed in the GFC1 and GFC2 models underestimates the magnitude of the hot gas' specific angular momentum. Consequently, the GFC1 and GFC2 models tend to underestimate the angular momentum of the accreted gas. 

As already mentioned, the new cooling model gives a specific angular momentum distribution (red solid line in top panel of Fig.~\ref{fig:slow_cool_j_distri}) that is lower than that in the simulation, with an offset about $0.1-0.2\,{\rm dex}$. This offset causes the new cooling model to underestimate the cooled-down angular momentum, but this underestimation is milder than in the GFC1 and GFC2 models. As discussed previously, this offset is mainly caused by the hot gas density profile used in the \GALFORM models, and if it were replaced with a more accurate profile, the underestimation would be reduced.

The distributions in the top panel of Fig.~\ref{fig:slow_cool_j_distri} are normalized by the magnitude of the mean specific angular momentum of the entire dark matter halo, $|\bar{\boldsymbol{j}}_{\rm dark}|=|\boldsymbol{J}_{\rm halo}|/M_{\rm halo}$. This figure shows that in the simulation, $|\boldsymbol{j}_{\rm hot}(r)|=|\bar{\boldsymbol{j}}_{\rm dark}|$ at $r\sim 0.3r_{\rm vir}$. As discussed above, the gas accreted between snapshots should come from $r<0.25r_{\rm vir}$, so estimating the specific angular momentum of the accreted gas as $|\bar{\boldsymbol{j}}_{\rm dark}|$, as done in the \lgalaxy model, tends to overestimate angular momentum in the hot gas halo regime.
 
Our analysis here implies that the gas cooled down between snapshots has specific angular momentum, $\boldsymbol{j}_{\rm cooling}$, lower than that of the dark matter halo. This seems opposite to the conclusion of \citet{Stevens_cooling_2017}, who state that, on average, $|\boldsymbol{j}_{\rm cooling}|\approx 1.4 |\bar{\boldsymbol{j}}_{\rm halo}|$. The \citeauthor{Stevens_cooling_2017} result is based on the \eagle simulation, which includes strong effects from feedback, unlike our simulation, which has feedback turned off. The differences between this work and \citeauthor{Stevens_cooling_2017} might therefore be caused by feedback processes. We postpone detailed investigation of this point to a future work.

\subsection{Comparison with different SA models}
\label{sec:comparison}
We now we compare the cumulative angular momenta of cooled-down gas measured from the hydrodynamical simulation following the method described in \S\ref{sec:method_j_measurement} with the predictions of different SA models.

For this comparison, we divide the haloes in the simulation into four samples according to their $z=0$ descendant masses, covering the ranges: 
\begin{itemize}
\item $10^{11}\Msol\leq M_{\rm halo}(z=0)< 3\times 10^{11}\Msol$
\item $3\times 10^{11}\Msol\leq M_{\rm halo}(z=0)< 10^{12}\Msol$
\item $10^{12}\Msol\leq M_{\rm halo}(z=0)< 10^{13}\Msol$
\item $M_{\rm halo}(z=0)\geq 10^{13}\Msol$
\end{itemize}
Haloes in the first mass range are mainly in the filamentary accretion regime, from high redshifts to $z=0$. Haloes in the third and fourth mass ranges go into the hot gas halo regime at low redshift, while the second mass range is a transition region. There are $1086$, $462$, $200$ and $24$ haloes in the four respective mass ranges.

\begin{figure*}
\centering
\includegraphics[width=1.0\textwidth]{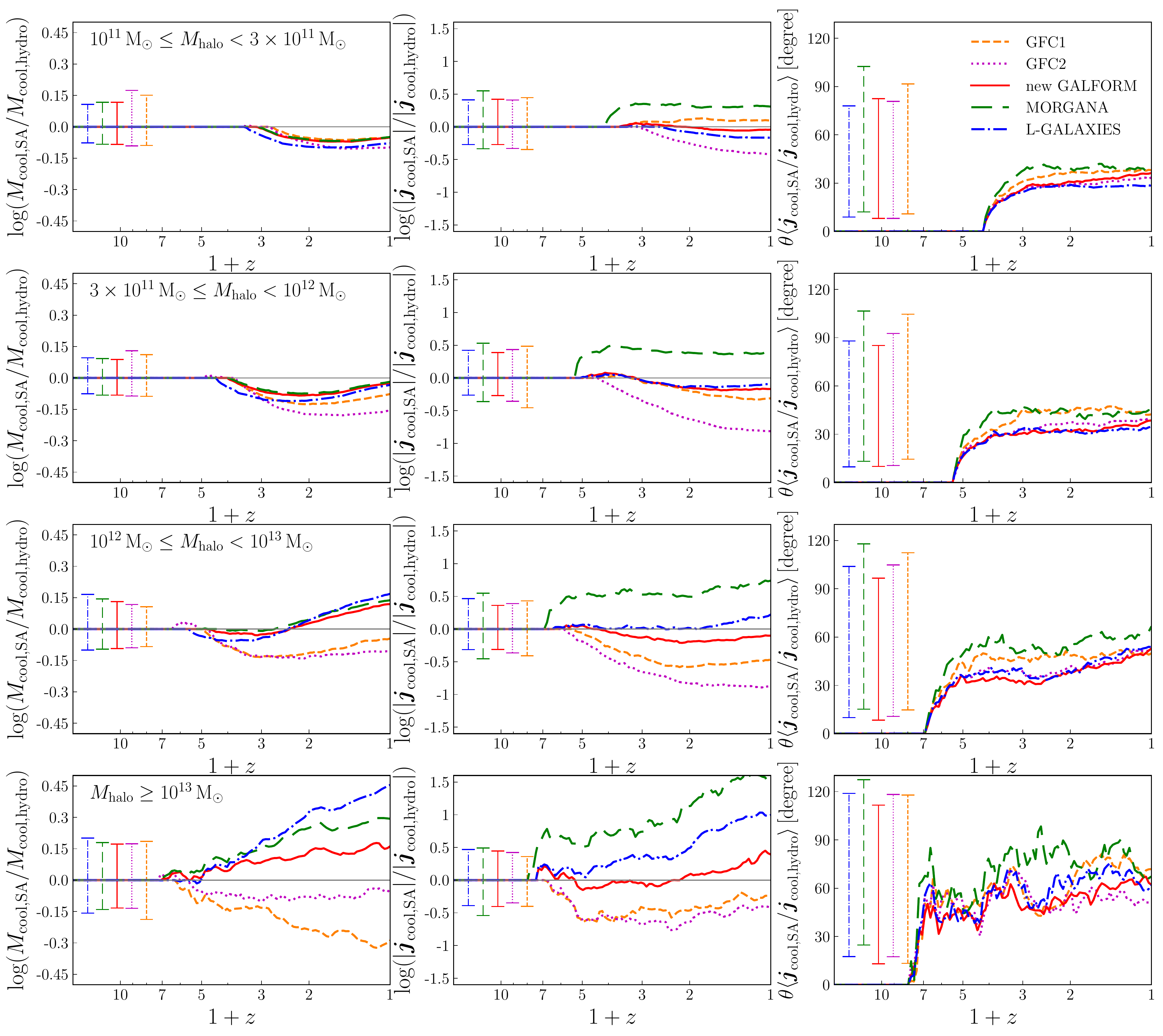}
\caption{Statistical comparison between the predictions of 5 different SA cooling models (\GALFORM GFC1, GFC2 and new cooling models, \lgalaxy, and \MORGANA)  and the hydrodynamical simulation. Each row corresponds to a halo sample selected on $z=0$ descendant masses, with the corresponding mass range given in the left panel in each row. Each column compares the SA predictions with the simulation results for a different quantity, with the lines showing the median difference between the SA model prediction and the simulation, plotted against redshift. The left column compares the cumulative mass of cooled-down gas, the middle column compares the magnitude of the mean specific angular momentum of cooled-down gas, and the right column compares the direction of the angular momentum of cooled-down gas. Different line styles are for different SA models, with the corresponding model names given in the key in the top right panel. The error bars with the same line styles show the typical $10-90$ percentile scatter for each case. }
\label{fig:cool_vJ_comp}
\end{figure*}

Fig.~\ref{fig:cool_vJ_comp} shows the medians and scatter of the individual halo differences between the SA models and the simulation, for the four above-mentioned halo samples. The left column shows the differences in the cumulative cooled-down mass, the middle column shows the differences in the magnitude of the specific cumulative angular momentum (defined in equation~\ref{eq:j_cool_def}), and the right column shows the direction offset between the specific angular momenta in the SA models and the simulation. Note that the vertical axis scale for cumulative mass is different from that for specific angular momentum. The differences in cooled-down masses predicted by various SA models and the simulation are smaller than those in specific cooled-down angular momenta. Therefore, the latter must be mainly caused by the modelling of angular momentum accretion discussed in \S\ref{sec:method_SA_j_model}, rather than the modelling of mass accretion. Below we discuss the results for each of the five SA models in turn.

The \MORGANA model assumes that the specific cumulative angular momentum of the galaxy is always equal to the mean specific angular momentum of the dark matter halo. This estimate would be accurate if the baryonic matter accreted onto a dark matter halo joined the central galaxy in the halo immediately, because the mean specific angular momentum of all baryons in a halo is similar to that of the dark matter halo, as shown in Fig.~\ref{fig:j_whole_halo_fast_cool}. 
The dark matter joins the halo as soon as it crosses the halo virial radius, $r_{\rm vir}$. However, for the baryonic matter there is a time delay between crossing $r_{\rm vir}$ and being accreted onto the central galaxy. Therefore, at any given time, there is some newly accreted baryonic matter in the halo that has not had enough time to reach the central galaxy. As argued below, if this newly accreted matter were added to the central galaxy then it would tend to increase the specific angular momentum of the latter. Therefore the \MORGANA model typically overestimates the specific cumulative angular momentum of the galaxy, as seen clearly in Fig.~\ref{fig:cool_vJ_comp}.

In the filamentary accretion regime, the accreted gas reaches the central galaxy within 1--2 halo dynamical times, $t_{\rm dyn}$ \citep{Danovich_j_2015}, where $t_{\rm dyn}=r_{\rm vir}/V_{\rm vir}$, with $V_{\rm vir}$ the halo virial velocity. This time duration is not very long, so the overestimation of the galaxy specific angular momentum in this regime is modest. In the hot gas halo regime, the accretion onto central galaxies is mainly determined by the gas cooling timescale, which can be much longer than $t_{\rm dyn}$, so the overestimation becomes worse.

We can examine this effect in a little more detail. Due to the time delay between gas crossing $r_{\rm vir}$ and being accreted onto the central galaxy, the specific angular momentum of the galaxy $|\boldsymbol{j}_{\rm gal}|\approx |\bar{\boldsymbol{j}}_{\rm halo,pre}|$, where $\bar{\boldsymbol{j}}_{\rm halo,pre}$ is the halo specific angular momentum at the time the most recently accreted gas crossed $r_{\rm vir}$. Then considering that the halo spin parameter, $\lambda \sim |\bar{\boldsymbol{j}}_{\rm halo}|/(\sqrt{2}r_{\rm vir}V_{\rm vir})$, has a  mean, $\bar{\lambda}$, that is almost independent of halo mass and redshift \citep[e.g.][]{cole2000,Bett_2007}, then one has on average $|\bar{\boldsymbol{j}}_{\rm halo}|\sim \sqrt{2}\bar{\lambda}r_{\rm vir}V_{\rm vir}$. This implies that on average $|\bar{\boldsymbol{j}}_{\rm halo}|$ increases with time, because with halo growth $r_{\rm vir}V_{\rm vir}$ increases with time. Therefore on average $|\bar{\boldsymbol{j}}_{\rm halo}| > |\bar{\boldsymbol{j}}_{\rm halo,pre}|$, and so if the gas newly accreted onto the halo were to be added to the central galaxy, it would shift $|\boldsymbol{j}_{\rm gal}|$ from $|\bar{\boldsymbol{j}}_{\rm halo,pre}|$ to $|\bar{\boldsymbol{j}}_{\rm halo}|$, and so tend to increase $|\boldsymbol{j}_{\rm gal}|$.

The \lgalaxy model instead assumes that the gas accreted onto a central galaxy in one timestep has specific angular momentum equal to $\bar{\boldsymbol{j}}_{\rm halo}$. As shown in \S\ref{sec:assumption_check_filament_regime}, this assumption is in reasonable agreement with our simulation (median magnitude ratio around $1.5$ and median angular offset about $40^{\circ}$) in the filamentary accretion regime, but \S\ref{sec:assumption_check_hot_halo_regime} implies that in the hot gas halo regime this assumption tends to overestimate the specific angular momentum of the gas accreted onto central galaxies. Haloes with higher $z=0$ descendant masses spend longer in the hot gas halo regime, and therefore the \lgalaxy model overestimates the specific cumulative angular momentum for these haloes more. This is what is seen in Fig.~\ref{fig:cool_vJ_comp}. 

As described in \S\ref{sec:lgalaxies_model}, the \SHARK model \citep{SHARK_model_2018} assumes that the magnitude of the mean specific angular momentum of the gas accreted in one timestep is equal to that of the dark matter halo at that time, but also that the accreted angular momenta at different timesteps all have the same direction. The assumption about the magnitude of the angular momentum change is the same as in the \lgalaxy model, but the assumption in \SHARK of a constant direction for the galaxy angular momentum replaces the vector addition in \lgalaxy with scalar addition. In the \lgalaxy model, the direction of the angular momentum accreted in one timestep is the same as $\bar{\boldsymbol{j}}_{\rm halo}$, and the latter changes its direction during cosmic evolution. Given two vectors of different directions, vector addition results in a smaller magnitude than scalar addition. Therefore, the \SHARK model has a higher risk of overestimating the cooled-down angular momentum in the hot halo regime than the \lgalaxy model. Note that because the \SHARK model mainly uses halo merger trees from N-body simulations, this risk could be reduced by using the full vector $\bar{\boldsymbol{j}}_{\rm halo}$ measured from the simulation, at least for well resolved haloes (having more than $300$ particles).

The GFC1 and GFC2 models do not consider in detail the evolution of the hot gas halo specific angular momentum profile, $\boldsymbol{j}_{\rm hot}(r)$, but simply assume that $\boldsymbol{j}_{\rm hot}(r) \propto r$. As shown in Fig.~\ref{fig:slow_cool_j_distri}, for haloes in the hot gas halo regime, this assumption tends to underestimate the specific angular momentum of the gas accreted onto central galaxies. This underestimation can be clearly seen from the lower two panels in the middle column of Fig.~\ref{fig:cool_vJ_comp}, which show the results for haloes with $M_{\rm halo}(z=0)>10^{12}\Msol$ (which are mainly in the hot gas halo regime). The underestimation in the GFC2 model is worse than in the GFC1 model, because the former includes the hot gas halo contraction at every timestep, and shells of hot gas are continuously moved to smaller radii during cooling, while in the latter this contraction is only included during the reset of the hot gas halo at each halo formation event. Smaller radii for the gas shells that are cooling then results in lower specific angular momentum, due to the assumption that $\boldsymbol{j}_{\rm hot}(r) \propto r$. This difference can  be clearly seen from the above-mentioned two panels of Fig.~\ref{fig:cool_vJ_comp}, as the magenta lines (for GFC2) are below the orange lines (for GFC1).

For haloes in the filamentary accretion regime, SA models typically still assume the existence of a hot gas halo, so the above-mentioned features of the GFC1 and GFC2 models also exist in the filamentary accretion regime. However, in this regime, SA models typically predict very fast cooling and free-fall limited accretion onto central galaxies. This means that all gas in the assumed hot gas halo would join the associated central galaxy quickly, and therefore the total angular momentum gained by the central galaxy is less sensitive to the assumed form of $\boldsymbol{j}_{\rm hot}(r)$ than in the hot gas halo regime. As a result, the underestimation in these two models is reduced for haloes that are mainly in the filamentary accretion regime, as can be seen from the upper two panels of the middle column of Fig.~\ref{fig:cool_vJ_comp}. Note however that the underestimation for the GFC2 model is still relatively strong in this regime.

The underestimation of the specific angular momentum in the GFC2 model should be reduced if a fixed form for $\boldsymbol{j}_{\rm hot}[M(<r)]$ (where $M(<r)$ is the total baryonic mass enclosed in radius $r$, normalized to $M(<r_{\rm vir})$), were adopted instead of a fixed form for $\boldsymbol{j}_{\rm hot}(r)$, such as $\boldsymbol{j}_{\rm hot}(r) \propto r$. Adopting $\boldsymbol{j}_{\rm hot}[M(<r)]$ was proposed in the original paper on the GFC2 model \citep{benson_bower_2010_cooling}, typically with $|\boldsymbol{j}_{\rm hot}|$ an increasing function of $M(<r)$. However, this is not likely to fully remove the underestimation. The reason is that adding new gas to the hot gas halo tends to reduce the normalized $M(<r)$ for any given gas shell, which, according to the fixed functional form  $\boldsymbol{j}_{\rm hot}[M(<r)]$, leads to lower $|\boldsymbol{j}_{\rm hot}|$ for that shell. 

The new \GALFORM cooling model models the accretion of angular momentum in more detail than the four earlier SA models discussed above. The middle column of Fig.~\ref{fig:cool_vJ_comp} shows that 
this model predicts magnitudes for the specific angular momentum of cooled-down gas that are overall in better agreement with our simulations than the other SA models, although the \lgalaxy model gives better agreement in the range $10^{12}\Msol\leq M_{\rm halo}(z=0)< 10^{13}\Msol$.
In general, the new cooling model also gives the best predictions for the direction of the angular momentum. The right column of Fig.~\ref{fig:cool_vJ_comp} shows that the median direction offset is between $30^{\circ}$ and $40^{\circ}$ for haloes with $M_{\rm halo}(z=0)< 10^{13}\Msol$, while the median is between $40^{\circ}$ and $60^{\circ}$ for haloes with $M_{\rm halo}(z=0)\geq 10^{13}\Msol$. For comparison, the median for directions chosen randomly over the range  $[0^{\circ},180^{\circ}]$ would be $90^{\circ}$. Note however that there is huge scatter in offset angle around the medians, and this is also true for the predictions of the other SA models considered in this work.

\begin{figure}
\includegraphics[width=0.45\textwidth]{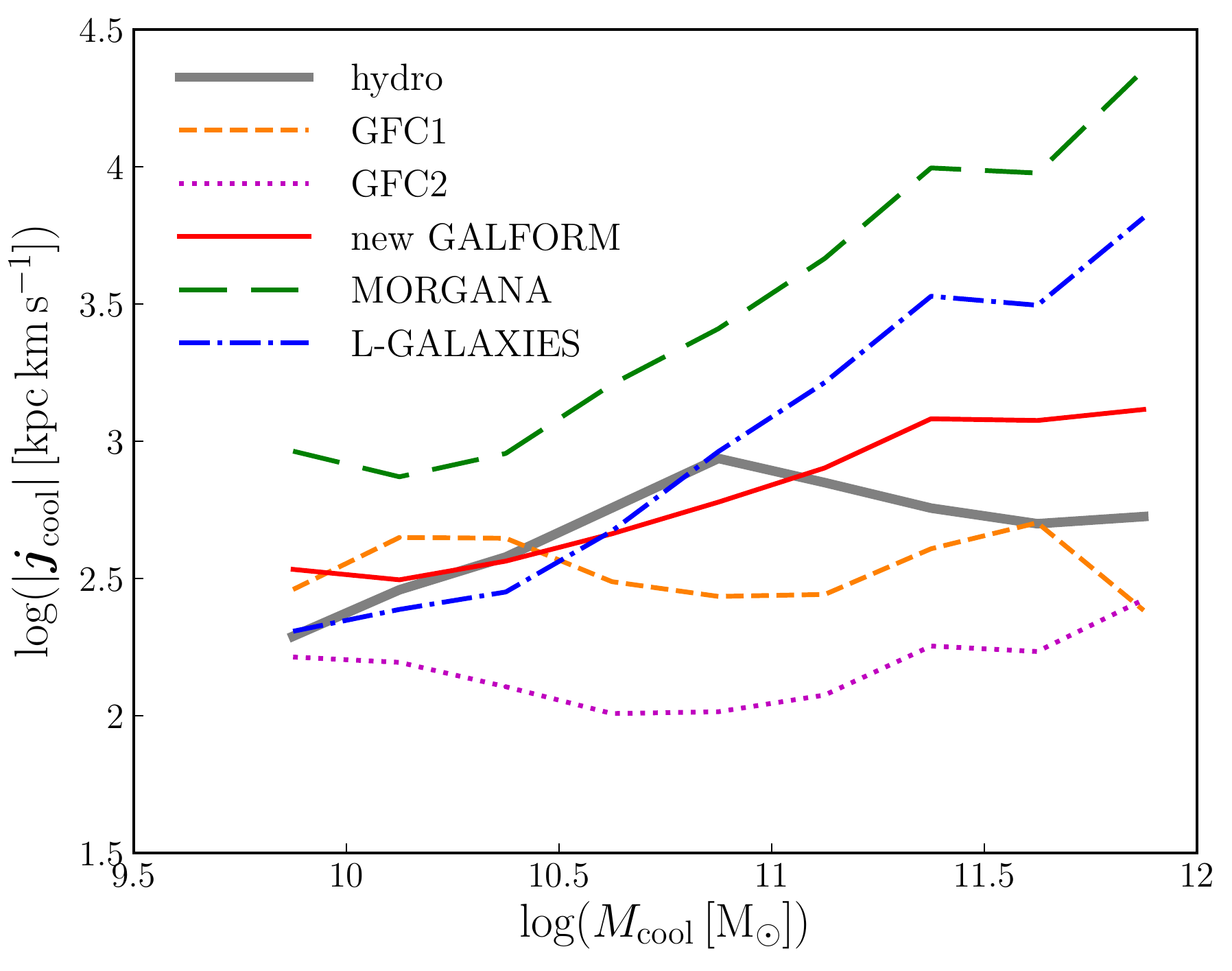}
\caption{Median relation between the specific angular momentum and mass of cooled-down gas at $z=0$. The gray thick solid line is for results extracted from the hydrodynamical simulation, while the other lines show the predictions of the various SA models considered in this work, as indicated in the key.}
\label{fig:j_mcool_correlation}
\end{figure}

The relation between the specific angular momentum and mass of cooled-down gas is also of some interest, since it is related to the angular momentum vs mass relation for galaxies \citep[e.g.][]{Fall2018}. We compare the median  relation at $z=0$ predicted by the hydrodynamical simulation and the SA models in Fig.~\ref{fig:j_mcool_correlation}. As shown in Fig.~\ref{fig:cool_vJ_comp}, the differences in the cooled-down masses predicted by the SA models and the simulation are smaller than those in the specific cooled-down angular momentum. Differences seen in Fig.~\ref{fig:j_mcool_correlation} are therefore mainly due to the latter. As discussed above, the \MORGANA model tends to overestimate the specific cooled-down angular momentum over the whole halo mass range considered here ($M_{\rm halo}(z=0) \geq 10^{11}\Msol$), so its prediction lies above that from the hydrodynamical simulation for the whole range of $M_{\rm cool}$. The old \GALFORM model GFC2 underestimates the specific cooled-down angular momentum, so its prediction lies below the simulation results for almost all values of $M_{\rm cool}$. The other three models, namely \lgalaxy, old \GALFORM model GFC1 and the new \GALFORM model, show deviations of the specific cooled-down angular momentum from the simulation results mainly for more massive haloes with $M_{\rm halo}(z=0)\geq 10^{12}\Msol$, which leads to the deviations seen in Fig.~\ref{fig:j_mcool_correlation} for high $M_{\rm cool}$. The GFC1 model tends to underestimate the specific cooled-down angular momentum, so it predicts a relation below that of the simulation for large $M_{\rm cool}$. On the other hand, both the \lgalaxy and the new \GALFORM model overestimate the specific angular momentum in massive haloes, so their predictions are above that of the simulation for large $M_{\rm cool}$. However, the overestimation in the new \GALFORM model is less than in the \lgalaxy model, so for large $M_{\rm cool}$ the new \GALFORM model is closer than \lgalaxy to the simulation results.

\section{conclusions and summary} 
\label{sec:summary} 
In this work we have compared the accretion of angular momentum onto galaxies due to cooling and accretion of gas from galaxy haloes as predicted by several different semi-analytic (SA) galaxy formation models with that measured in a hydrodynamical simulation of galaxy formation performed using the state-of-the-art moving-mesh code \AREPO.  Both the SA models and the simulation are run without any feedback or metal enrichment, in order to focus on the angular momentum accretion associated with gas cooling. The SA models considered here are the \MORGANA model, the \lgalaxy model and the \GALFORM model. For \GALFORM we considered two older cooling models GFC1 \citep{galform_bower2006} and GFC2 \citep{benson_bower_2010_cooling} and the new cooling model \citep{new_cool}.

These selected SA models cover a wide range of sophistication in the modelling of angular momentum accretion. The \MORGANA model does not follow the angular momentum flow associated with the mass flow due to cooling and collapse of gas in the halo, but simply assumes that at any given time, the mean specific angular momentum of a central galaxy in a halo equals that of its host halo. The \lgalaxy model tries to follow the angular momentum flow by assuming that the mean specific angular momentum associated with the gas flow onto a central galaxy at a given time is the same as that of the host dark matter halo at that time. The \GALFORM models calculate the angular momentum of the cooling mass flow by modelling the specific angular momentum distribution, $\boldsymbol{j}_{\rm hot}(r)$, of the hot gas halo, from which the cooling flow originates. The GFC1 and GFC2 models assume that $\boldsymbol{j}_{\rm hot}(r)\propto r$, which is inspired by hydrodynamical simulations without cooling, while the new \GALFORM cooling model further considers the evolution of $\boldsymbol{j}_{\rm hot}(r)$ induced by cooling in the halo, under the assumption that each Lagrangian hot gas shell conserves its angular momentum.

We first check various assumptions involved in the SA calculations. In the simulation, haloes at high redshifts and less massive than $3\times 10^{11}\Msol$ accrete gas mainly through cold filaments. The accreted gas reaches the central galaxies through an approximately free-fall collapse, and we found that at least on average this gas approximately conserves its angular momentum during the infall from $r=r_{\rm vir}$ to $r\sim 0.1r_{\rm vir}$, with $r_{\rm vir}$ being the halo virial radius. This is consistent with previous works, e.g. \citet{Danovich_j_2015}. \citeauthor{Danovich_j_2015} further pointed out that within $r\sim 0.1r_{\rm vir}$, infalling gas loses angular momentum due to gravitational torques from the central galaxy; however, these torques should not alter the total change in angular momentum of the central galaxy due to gas accretion. Therefore, assuming that the gas conserves its angular momentum during its motion from $r_{\rm vir}$ onto the central galaxy and delivers this angular momentum to the latter seems to be approximately valid in this filamentary accretion regime. 

The mean specific angular momentum of the halo gas, which in this regime is dominated by cold filaments, is about two times higher than that of corresponding dark matter haloes. This is also consistent with previous works \citep[e.g.][]{Stewart_j_2013,Danovich_j_2015}. However, only the halo gas in the central region of a halo is about to be accreted onto the central galaxy in the current timestep. Therefore, it is the angular momentum of this gas that should be compared to the accreted angular momentum calculated in SA models. We checked that the specific angular momenta of accreted gas calculated in the \lgalaxy and new \GALFORM cooling model are in good agreement with the mean specific angular momentum of this halo gas for haloes in this regime.

The low-redshift, high mass haloes in the simulation form roughly spherical hot gas haloes. We checked that the median angular momentum profile, $\boldsymbol{j}_{\rm hot}(r)$, predicted by the new \GALFORM cooling model is in approximate agreement with that measured from the simulation (although with an offset in amplitude that is mainly due to the hot gas density profile used in the \GALFORM models differing from the simulation in the range $0.7r_{\rm vir}\lsim r\lsim r_{\rm vir}$). It seems that the combination of assumptions in this SA model for deriving $\boldsymbol{j}_{\rm hot}(r)$, including that each Lagrangian hot gas shell conserves its angular momentum during hot halo contraction, is roughly consistent with the hydrodynamical simulation.

We then assess the accuracy of these SA model predictions for accreted angular momentum by means of a statistical comparison of the mean specific angular momenta of the cooled-down gas accumulated in central galaxies, as predicted respectively by the SA models and the simulation. We compare both magnitudes and directions for this specific angular momentum.

We found that the \MORGANA model tends to overestimate the magnitude of this specific angular momentum. This is because the mean specific angular momentum of the host halo, which is used in the \MORGANA calculation, includes the contribution from recently accreted dark matter, but the corresponding recently accreted gas, which usually has high angular momentum, has not had enough time to reach the central galaxy.

We found that the \lgalaxy model tends to overestimate the angular momentum of accreted gas in the hot gas halo regime, i.e.\ for low-redshift, high mass haloes, because the specific angular momentum that it assumes for the cooling flow is higher than that measured in the simulation.

The GFC1 and GFC2 \GALFORM models tend to underestimate the angular momentum of gas accreted onto galaxies. This mainly stems from the assumption that $\boldsymbol{j}_{\rm hot}(r)\propto r$, which does not include the effect of contraction of the hot gas halo due to cooling. Compared to the $\boldsymbol{j}_{\rm hot}(r)$ measured from the simulation with cooling, this assumed form underestimates $\boldsymbol{j}_{\rm hot}(r)$ for the small radii from which the hot gas cools down and accretes onto the galaxy.

Comparing the various SA models, the new \GALFORM cooling model appears in best agreement with the simulation overall for the magnitude of the angular momentum accreted, although this does depend somewhat on the halo mass range. The median of the offset in direction between the angular momentum in the SA model and in the simulation is between $30^{\circ}$ and $60^{\circ}$, which we regard as a reasonable degree of alignment (although with a large scatter).

In summary, although SA models all adopt simplified assumptions, some of them are sophisticated enough to make fairly reliable predictions for the angular momentum brought into galaxies by gas accretion from the halo, at least in the simplified situation without feedback. This provides a cornerstone for the reliability of further SA predictions, such as galaxy sizes and star formation histories. A direct modelling of cold filamentary accretion in SA models that goes beyond assumptions of spherical symmetry may reduce the scatter seen in the current work, and thus further improve the predictions for angular momentum accretion onto galaxies.

When feedback is included, the reincorporation of gas ejected by feedback becomes another source of accretion of angular momentum. If the ejected gas does not move far from its previous host galaxy, and falls back to the galaxy quickly, then the chance for significant action by gravitational tidal torques is low, and the gas' angular momentum should be approximately unchanged. In this situation, adding feedback should not lead to results strongly different from those without feedback. On the other hand, if feedback processes eject gas outside of dark matter haloes, then during reincorporation onto haloes, the ejected gas probably feels tidal torques similar to those felt by the gas and dark matter accreted during structure growth, and therefore these two kinds of material gain similar specific angular momenta. Because on average halo specific angular momentum grows with time, the newly accreted material has higher specific angular momentum, so the specific angular momentum of the reincorporated gas should also be higher. Adding this reincorporated gas to central galaxies is expected to raise their specific angular momenta, and therefore to alleviate the overestimation of angular momentum in the \MORGANA and \lgalaxy models seen in this work. The exact level of this alleviation should depend on feedback strength and can only be revealed through a detailed comparison between hydrodynamical simulations and semi-analytic models that both include modelling of feedback.


\section*{Acknowledgements}
We thank Prof.\ Volker Springel for providing a specially modified version of the \AREPO code for our project. We also thank Prof.\ Volker Springel, Dr. R$\ddot{{\rm u}}$diger Pakmor, Dr. Sownak Bose and Dr. Yan Qu for their help in using \AREPO code. 
This work was supported by the Science and Technology Facilities Council [ST/L00075X/1 and ST/P000541/1], European Research Council (ERC) Advanced Investigator Grant DMIDAS [GA 786910] and Natural Science Foundation of Shanghai [20PJ1412300]. 
This work used the DiRAC Data Centric system at Durham University, operated by the Institute for Computational Cosmology on behalf of the STFC DiRAC HPC Facility (www.dirac.ac.uk). This equipment was funded by BIS National E-infrastructure capital grant ST/K00042X/1, STFC capital grant ST/H008519/1, and STFC DiRAC Operations grant ST/K003267/1 and Durham University. DiRAC is part of the National E-Infrastructure.

\section*{Data availability}
The data underlying this article will be shared upon reasonable request to the corresponding author.

\bibliographystyle{mn2e}

\bibliography{paper}


\appendix
\section{Effects of different aperture sizes on angular momentum measurement}\label{app:aperture_size}

\begin{figure*}
\centering
\includegraphics[width=1.0\textwidth]{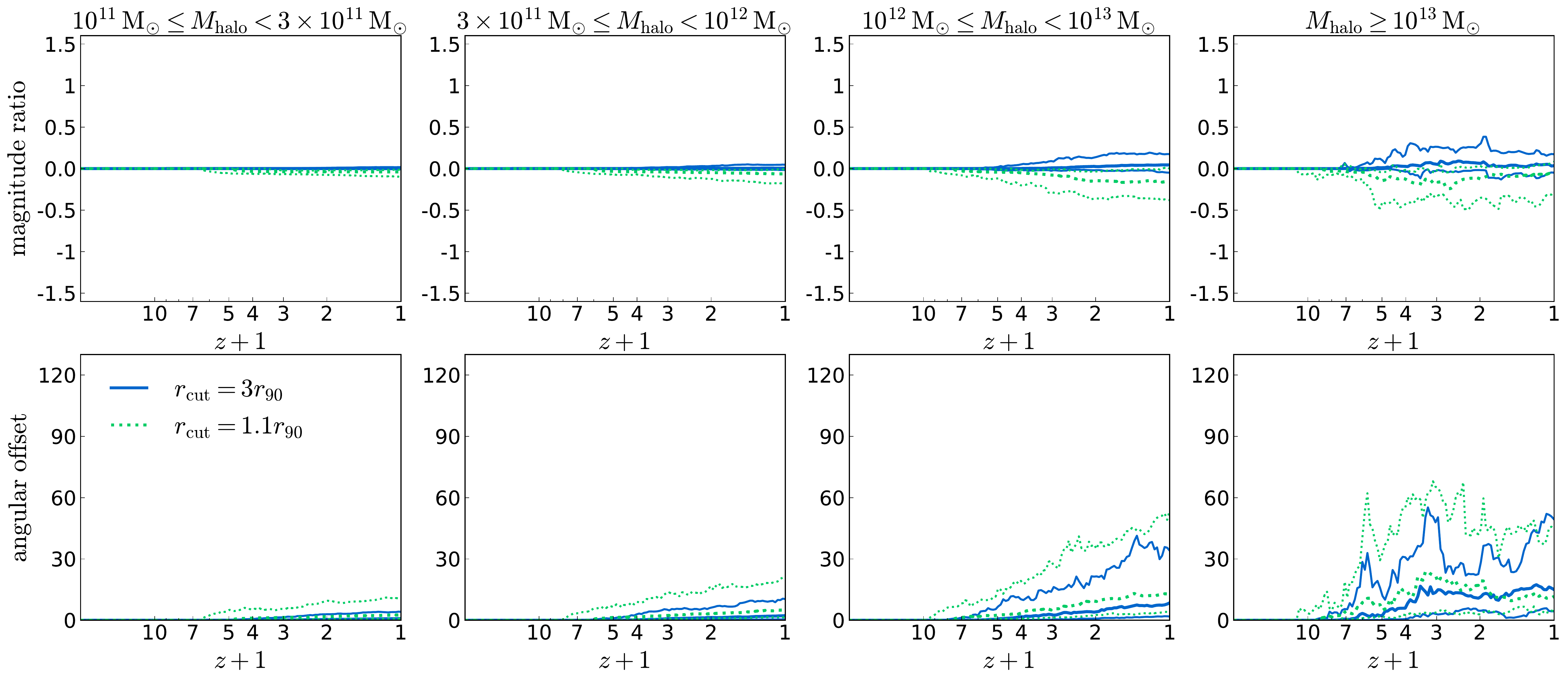}
\caption{Comparison of cumulative cooled-down angular momenta measured with different aperture sizes. The angular momenta measured with aperture radii $1.1r_{\rm 90}$ and $3r_{\rm 90}$ are compared with the default results, which use aperture radius $2r_{\rm 90}$. The top row compares the magnitudes of the angular momenta calculated with the two apertures, while the bottom row compares the directions. Each column is for a halo mass bin, with the halo mass range given at the top of the column. In each panel, the dotted and solid lines show the comparisons for aperture radii of $1.1r_{\rm 90}$, and $3r_{\rm 90}$ respectively. Thick lines show medians, while thin lines indicate 10-90 percentiles. The vertical axis scales are chosen to be the same as those for the middle and right columns of Fig.~\ref{fig:cool_vJ_comp} to make it easier to compare the differences caused by different aperture sizes and the differences between SA models and the simulation. This figure suggests that the former effect is smaller than the latter, so the latter comparison is not sensitive to the particular choice of aperture size.}
\label{fig:J_r_cut_comp}
\end{figure*}

Here we compare the cumulative cooled-down angular momenta measured with different aperture sizes. The default results are derived with the aperture size $2r_{\rm 90}$. Here, as stated in \S\ref{sec:method_j_measurement}, $r_{\rm 90}$ is derived by first selecting all stellar particles belonging to the most massive subgroup of a given Dhalo and lying within $50$ comoving kpc from the subgroup centre, and then taking the $90$ percentile of the distances of these particles to the subgroup centre. These default results are then used in the comparison with the predictions of the SA models in the main body of this paper. Here we also derive results with aperture sizes $1.1r_{\rm 90}$ and $3r_{\rm 90}$, and compare those with the results for the default aperture size.  Fig.~\ref{fig:J_r_cut_comp} shows the result of this comparison, and it shows that the differences in angular momentum magnitudes and directions caused by varying aperture size are smaller than those between the predictions of the SA models and the simulation (shown in Fig.~\ref{fig:cool_vJ_comp}). Therefore, the differences seen between the SA models and the simulation are not very sensitive to the choice of aperture size.

\section{Effects of subtracting gravitational torque contribution to angular momentum}\label{app:torque_subtraction}

\begin{figure*}
\centering
\includegraphics[width=1.0\textwidth]{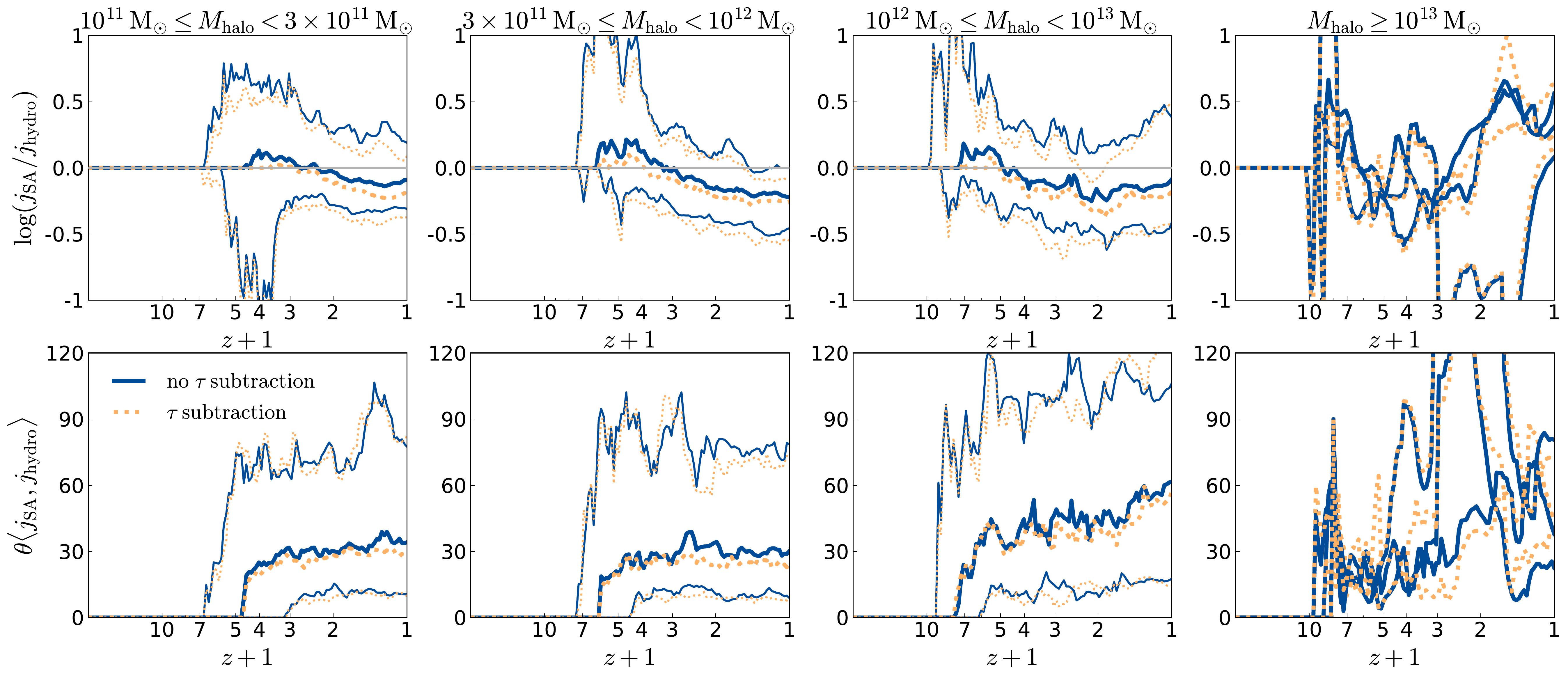}
\caption{A similar comparison to that in Fig.~\ref{fig:cool_vJ_comp}, but for simulation results derived respectively with and without subtracting the estimated effect of gravitational torques, for the $25\Mpc$ simulation cube. For simplicity, only the SA predictions for the new \GALFORM cooling model are shown. Solid lines are for results without the effects of torques subtracted, and dotted lines are for results with this subtraction. Each column is for a different halo mass range, with the mass ranges shown at the top of each column. The left three columns are for haloes with $M_{\rm halo}\leq 10^{13}\Msol$, and statistical results are shown, with thick lines for medians, and thin lines for the $10-90$ percentile range. The rightmost column is for haloes more massive than $10^{13}\Msol$. In the $25\Mpc$ simulation there are only three such haloes, so results of individual haloes are shown instead of medians and percentiles. }
\label{fig:dJ_torque_remove_comp}
\end{figure*}

Here compare two different methods to estimate the angular momentum delivered to central galaxies by gas accretion, using the simulation run in the $25\Mpc$ cube. The first method is the same as that in \S\ref{sec:method_j_measurement}, namely the angular momentum delivered between snapshots ${\rm i}$ and ${\rm i}+1$ is estimated as $\Delta\boldsymbol{J}_{\rm i+1}$. In the second method, this angular momentum is instead estimated as $\Delta\boldsymbol{J}_{\rm i+1}-\Delta\boldsymbol{J}_{\rm \tau}$, which includes an approximate correction for the effect of external gravitational torques represented by $\Delta\boldsymbol{J}_{\rm \tau}$ (calculated as described in \S\ref{sec:method_j_measurement}). These changes in angular momenta between snapshots are then added up and divided by the cumulative accreted mass to derive the specific cumulative angular momenta. These specific angular momenta from the simulation are then compared with those predicted by the new \GALFORM cooling model. The results are shown in Fig.~\ref{fig:dJ_torque_remove_comp}. They imply that the comparison between SA models and simulations is not sensitive to the subtraction of $\Delta\boldsymbol{J}_{\rm \tau}$. Therefore we have not made this correction to the results shown in the main body of the paper, which are calculated for the simulation run in the $50\Mpc$ cube.

\end{document}